\documentclass[12pt]{article}

\usepackage{amsfonts}
\usepackage{amsmath}
\usepackage{amssymb}
\usepackage{graphicx}

\def\be{\begin{equation}}
\def\ee{\end{equation}}
\def\bea{\begin{eqnarray}}
\def\eea{\end{eqnarray}}
\def\beann{\begin{eqnarray*}}
\def\eeann{\end{eqnarray*}}
\def\nn{\nonumber}
\newcommand\qed{\begin{flushright} $\Box$ \end{flushright}}

\newtheorem{rem}{Remark}

\newtheorem{lem}{Lemma}

\usepackage{epsfig}

\title{Operationalization of Relativistic Motion}

\author{Bruno Hartmann\footnote{brunohartmannjr@gmail.com}\\
\textit{Perimeter Institute for Theoretical Physics,}\\
\textit{Waterloo, ON N2L 2Y5, Canada,}\\
\textit{Humboldt University, D-12489 Berlin, Germany}}

\begin{document}

\maketitle

\textbf{Abstract:} We demonstrate the definition of basic observables from physical operations, the key to overcome hidden stumbling blocks and apparent paradoxes from unscrutinized (classical) formalisms. We develop Helmholtz program of basic measurements for relativistic motion. We define the basic observables by direct comparison: ''longer than'' if one object or process covers the other. To express the spatiotemporal order also numerically (how many times longer) we cover them by a locally regular grid of light clocks. These are the basic physical operations. From their interrelation we derive mathematical relations, e.g. for different observers the formal Lorentz transformation; for accelerating observers we reveal a measurement-methodical view on the apparent Twin paradox.
\\

One usually explains kinematics axiomatically. So one can trace back the whole mathematical formalism to a manageable system of initial propositions which are logically independent from one another. Though this formal bookkeeping of physics already begins in the abstract. The axiomatic formulation assumes scalars, four-vectors, metric etc. as known objects of its description. Without further implicit assumptions it lacks interpretation and physical meaning. Origin, scope and limitations of the variables and algebra remain unclear. The lack of alternative approaches seems to justify the formal path for developing novel theories. For the foundation of elementary kinematics (next also for dynamics \cite{Hartmann-KM_Dyn}) we develop a complementary program; we begin from the primary measurement operations \{\ref{Kap - SRT Massbestimmung}\}.

Our objective is a foundation of relativistic kinematics from the operationalization of its basic observables. The goal is not to change or improve the mathematical structure, but to gain a deeper physical understanding of kinematics. Like Einstein \cite{Einstein-Grundlagen der ART} for the concept of simultaneity we reveal the underlying physical operations. We begin from undisputed natural and measurement principles. We stress the active role of physicists, the interventions to define basic observables, quantification and then derive the four-vector formulation second. With Helmholtz' method (general discussion in retrospect \{\ref{Kap - math formulation of physical operations}\}) we will show, how physics generates its own mathematics in empirical practice.

Theodor H\"ansch - inventor of the optical frequency comb generator which facilitates the construction of most precise clocks - defines: ''Time is what one measures with a clock''. In his case a light clock. The origin of basic \emph{reference devices} and measurement \emph{procedures} is not in the domain of non-empirical mathematics. We require them (like Einstein's clock postulate and laser ranging technique) \emph{before} having a theory of matter and as a basis for developing the latter \cite{Hartmann-KM_Dyn}. We develop all protophysical prerequisites from everyday work experience. We make a digression into watchmaking and understand by what actions one provides these reference devices if one did not have them before \{\ref{Kap - SRT Massbestimmung - classical metric - Watchmaking}\}. With classical rulers and clocks one determines the universal motion of light \{\ref{Kap - SRT Massbestimmung - light principle}\}, it propagates locally uniform. For basic measurements we introduce (light) clocks as an unstructured unit \{\ref{Kap - SRT Massbestimmung - light clock}\}.

Let every observer place them side by side and one after another until the measurement object is covered; for the technical description we introduce \emph{measurement termini} (simultaneity lines, projections etc.). In the mathematical formulation of the procedures all corresponding \emph{terms} have physical meaning. From the underlying operations we derive the Lorentz symmetry \{\ref{Kap - Masszsh}\}. Every formal calculation, e.g. in the configuration of the apparent Twin paradox \{\ref{Kap - Twinparadox}\}, assumes connected basic measurement operations. For an accelerating observer they become impracticable. From vivid pre-theoretic principles we develop all mathematical variables and operations and finally the relativistic equations.

%%%%%%%%%%%%%%%%%%%%%%%%%%%%%%%%%%%%%%%%%%%%%%%%%%%%%%%%%%%%%%%%%%%%%%%%%

\section{Measurement operations}\label{Kap - SRT Massbestimmung}

For the origin of colloquial notions - \emph{motion}, \emph{space} and \emph{rigid body} - from common sensual experience we refer to Poincare \cite{Poincare - Wissenschaft und Hypothese} and Mach \cite{Mach - Raum und Geometrie}. According to Poincare \emph{geometric properties} essentially characterize the relative \emph{motion} between neighboring \emph{objects}. Leibniz characterizes ''space'' and ''time'' as relations between the observable things. \emph{Space} brings order into things which happen simultaneously. \emph{Time} brings order into things which happen sequentially. From everyday practice one knows the direct comparison
\begin{itemize}
\item   $>_{l}$ \;\;\; if two extended objects lie on top of each other - one will \emph{cover} the other
\item   $>_{t}$ \;\;\; if two processes begin simultaneously - one will \emph{outlast} the other.
\end{itemize}
The \emph{ordering relation} is reproducible in an observer independent way. Next one wants to find ''how many times'' longer.

For reproducible measurements one provides sufficiently constant reference devices and standardized procedures. The \emph{construction} and the \emph{conventions} historically developed from daily work experience; we sketch the transition to physical experimentation. One works with natural objects in a natural environment. Their behavior depends on external conditions (some known and others undiscovered). One wants to control the interrelation of work conditions; it pays off. We regard the origin of basic measurements as a standardization in the conduct of reproducible experiments (to specify known work conditions \cite{Ruben - Arbeitskonzept}). With empirical knowledge on feasibility and outcomes of work \emph{actions} one can probe the objects and rehearse expedient ways of handling. We define all basic observables (length, duration; in \cite{Hartmann-KM_Dyn} also impulse, inertial mass, capability to work/energy) and the associated comparison and concatenation operations in the practical domain. The development has a social dimension: a master inherits the demonstrable practice to a student, first simply by pointing a finger and then defines a colloquial and technical language. We presuppose usage of common denominations (''$\mathcal{A}$lice, $\mathcal{B}$ob and $\mathcal{O}$tto move relative to one another.'', ''They signal with light.'' etc.) with their common meaning as a known part of work experience.\footnote{We explain the meaning of colloquial expressions by exemplary demonstration (of sufficiently constant phenomena). We can neither demonstrate pure matter in isolation from its behavior nor pure behavior detached from matter. In colloquial speech we express a demonstrable fact by a simple descriptive sentence like ''Otto is long''. These represent the smallest unit of meaning. The subject Otto $\mathcal{O}$ and its attribute length $l$ are distinguishable but inseparably unified \cite{Peter '76 - Praedikationstheorie} \cite{Vati - Logik und Arbeit}. The subject terminus ''long Otto'' emphasizes the subject $\mathcal{O}$ which embodies the property long $l$; we symbolize the \emph{long object} by $\mathcal{O}_l$. The predicate terminus ''Otto's length'' emphasizes the property which Otto represents; we symbolize the \emph{object's length} $l_{\mathcal{O}}$. From elemental operations on tangible things $\mathcal{O}_l$ we develop basic measurements for the attribute $l_\mathcal{O}$.} With Lorenzen, Janich \cite{Janich Das Mass der Dinge} we can presuppose (circularity free and without mathematical presuppositions) that every observer can manufacture ''straight'' ''rigid'' rulers and ''uniform running'' clocks \{\ref{Kap - SRT Massbestimmung - classical metric - Watchmaking}\}. In a direct measurement one \emph{concatenates} ''$\ast_{s}$'' the rulers ''$\mathcal{R}$'' side by side in a straight way until the layout, symbolized $ \mathcal{R} \ast_{s} \ldots \ast_{s} \mathcal{R} \sim_{s} \mathcal{O}$, covers the object. The ordering relation ''longer than'' becomes measurable $s_{\mathcal{O}} = \sharp\left\{\mathcal{R}\right\} \cdot s_{\mathcal{R}}$ by the number of connected rulers and their standard length $s_{\mathcal{R}}$; similarly for durations.

In starting figure \ref{figure-1}a we illustrate objects and observers in motion. Consider a (hidden) railway track along which $\mathcal{A}$lice, $\mathcal{B}$ob and $\mathcal{O}$tto specify their relative motion and the light. After including the historical depth of work experience they are equipped with the local Euclidean metric. We will demonstrate the transition to relativistic kinematics. Each observer measures $\mathcal{O}$tto's relative motion with classically constructed light clocks. They place them one after the other or side by side; connected by coinciding rays of light \{\ref{Kap - SRT Massbestimmung - direct connecting products}\}. A regular grid of light clocks covers their relative distances. Each building block is congruent with the next; by counting them they measure the magnitude of the length. They measure their relative motion with (the motion of light in) their reference device.

\subsection{Watchmaking}\label{Kap - SRT Massbestimmung - classical metric - Watchmaking}

The protophysical foundation of Euclidean geometry \cite{Janich Das Mass der Dinge} explains the standardization of length comparisons circularity free in the categories of purpose and expedient means of everyday work. One works on raw materials and reshapes them for practical needs. One builds rulers and clocks as sufficiently constant representatives of ''length'' and ''duration''. The success of (tentative) manufacturing methods is secured by \emph{test procedures} for the \emph{straight form} of a constructed ruler and the \emph{uniform running} of a clock \cite{Janich Das Mass der Dinge}.\footnote{Before $\mathcal{B}$ob can specify the form of $\mathcal{O}$tto's relative motion he has to find out if his own clock provides a uniform ticking reference. Before $\mathcal{A}$lice can determine whether $\mathcal{O}$tto's nose is crooked she needs to know if her ruler is straight. The test rules for admissible reference devices do \emph{not} presuppose an already existent \emph{prototype} for a straight line and an ur-clock which one can simply copy or transport.} For this reason \emph{measurement instruments} are to be understood - not simply as arbitrary designations of natural objects but - as \emph{artifacts}; our manufacturing actions must realize test norms \cite{Schlaudt}.

The testing method for geometrical shapes originates from grinding practice: A body has a ''flat'' surface if one can produce two moldings of that body and then fittingly (!) shift the two imprint surfaces against one another. If there is a gap continue grinding them against one another; make two new moldings and check again. Similarly if one has manufactured a body with two flat surfaces which intersect one another, then the intersecting edge represents a ''straight'' line. The test norms originate from intuitively controlled actions in everyday (technical) work, which is governed by the rationality of purpose and expedient means. Ultimately one explicates expedient work norms as measurement norms. Practicable rules of pre-scientific technical behavior develop into norms for measurement operations \cite{Schlaudt}.

A watchmaker evaluates by test procedures whether a tentative device runs uniform. In the empirical interplay of analyzing manufacturing conditions and examining the respective products the \emph{manufacturing method} is continually refined until the device \emph{realizes} the aspired \emph{ideal} of uniform motion sufficiently precise. In this process we make the practical experience that the ideal is never completely realizable. The closer one wants to approach the more effort and workload is required in the production and also for the conservation of the product (shielding fragile clock). He takes guidance by a test procedure: Take two structurally identical copies of the clock and align them such that their clock hands are running straight into fixed (e.g. perpendicular) directions. Then one can couple the motion of the two clock hands e.g. by a mechanical transmission; draw down the stretch of way of their superposed motion and \emph{check} its geometric form. The clocks run uniform if - independently from when each was started and coupled together - their superposed stretch of way always has the form of a straight line. As before the path in question represents the ideal of a straight line if any two segments can be fittingly (!) shifted against one another.\footnote{The protophysical norm for \emph{uniform} motion originates from a test of the \emph{straight} shape of rigid objects.}

A clock is manufactured and tested as a representative for a uniform motion. By metricizing the length of the traversed stretch of way (of the moving clock hand) one obtains a metrical measurement instrument for ''durations''. In practice (accumulated friction etc.) clocks will run (approximately) uniform for only finite durations. Such ''finite duration'' measurement standards can be aligned synchronously one after the other to cover longer processes. By this \emph{substitution} we measure the magnitude of ''durations''. We arrive at classical Galilei kinematics for space and time.
%[Vorausgesetzten Erzeugung klassischer Messstandards]
Despite the uniform motion of the clock hand - the clock (housing) can be under acceleration, sitting still or free falling.

\subsection{Principle of Inertia}

We link uniform motion to the behavior of natural (work) objects by the principle of inertia. Bodies move (without external agent) on their own. ''Every body with no (external) forces acting on it remains - as judged from the (inertial) lab - in a state of rest or of uniform rectilinear motion'' \cite{Sexl Urbantke Relativity}. In isolation their motion is preserved. We identify the presence of interactions by changing state of motion and the absence from practical reasoning. We postulate an inertial reference as a reproducible experimental prerequisite which we must shield from all external disturbances. We can provide it after probing all empirical conditions of an interaction (e.g. set up a billiard table \emph{horizontally} and \emph{test} that a prepared arrangement of object balls does not roll off to any side before the actual experiment). As Galilei and Huygens we develop via principle of inertia and relativity principle elementary dynamical concepts \emph{before} the latter were transferred by Newton onto gravitational systems. There the (initially practically solved) problem of selecting inertial references is revised; for building and steering machines (for the arising need to mechanize tool use based on the division of labor in industrial revolution) the latter had no practical significance.

Newton could draw on (in \emph{top}-heavy circles proscribed) ''literature of practical (\emph{hand}craft) mechanics on problems of machine construction and work economy, which is considered to little by historiography of science.'' For the context of origin of dynamical concepts Wolff's genetic reconstruction of Impetus theory - mechanics in epoch from 6. to 17. century - provides ''plausible arguments for the proposition that the inner conceptual content of mechanics was influenced by motives, which developed during that economic and technical revolution'' \cite{Wolff - Geschichte der Impetustheorie}.

%%%%%%%%%%%%%%%%%%%%%%%%%%%%%%%%%%%%%%%%%%%%%%%%%%%%%%%%%%%%%%%%%%%%%%%%%%%%%%%%%

\subsection{Light principle}\label{Kap - SRT Massbestimmung - light principle}

In order to give physical meaning to the concept of time Einstein \cite{Einstein-Grundlagen der ART} requires the use of some process which establishes relations between distant locations. In principle one could use any type of process. Most favorable for the theory one chooses a process about which we know something certain. For the free propagation of light this holds much more than for any other process.

\begin{figure}         %Bem.: height skaliert Dateibild auf Zielgroesse im Dokument
  \begin{center}         %eps Dateien sind einfuegbar und auch in dvi sichtbar
                         %pdf nur bei direkter pdf-Kompelierung sichtbar
  \includegraphics[height=7.3cm]{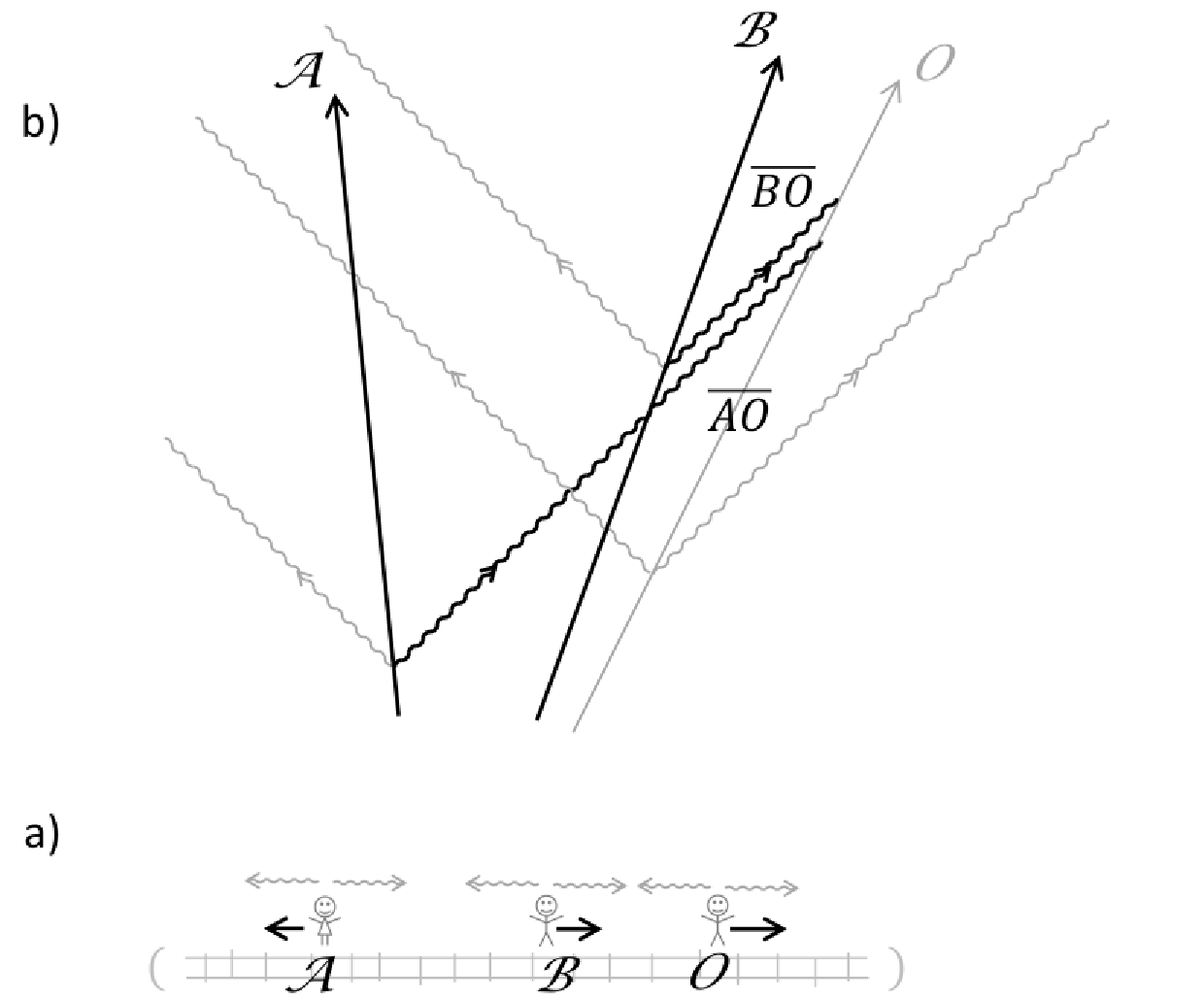}  %\\ if second picture aligned %below...
  \end{center}
  \vspace{-0.0cm}
  \caption{\label{figure-1} a) moving objects and moving light b) interrelation of corresponding processes
    }
  \end{figure}

One measures the motion of light with rulers and clocks. We depict the relative motion between all objects in a \emph{spacetime diagram} (see figure \ref{figure-1}b).
$\mathcal{A}$lice, $\mathcal{B}$ob or $\mathcal{O}$tto may move equally or not, but no object can overtake free light. Let $\mathcal{A}$lice and $\mathcal{B}$ob emit light towards $\mathcal{O}$tto. It propagates independently no matter how they move the source. If $\mathcal{A}$lice sends her light to $\mathcal{O}$tto and shortly after it passes $\mathcal{B}$ob he sends his own light to $\mathcal{O}$tto as well, then both rays $\overline{\mathcal{AO}}$ and $\overline{\mathcal{BO}}$ coincide. One measures the ''magnitude'' of distances and durations and the ''form'' of motion with classical rulers and clocks. They approximate a straight line and uniform motion. By local comparison with these reference instruments free light propagates in a uniform and straight way. In a spacetime diagram we represent it by a \emph{straight line}. Locally the light $\mathcal{A}$lice or $\mathcal{B}$ob send to $\mathcal{O}$tto $\overline{\mathcal{AO}}$ and $\overline{\mathcal{BO}}$ remains parallel.

With the classical metric (in the domain of classical measurements of length and duration) we discover: \emph{Locally} free light represents a \emph{uniform, isotropic and straight form of motion}. It provides a universal reference for any intrinsic observer. Based on the light principle we conduct laser ranging measurements. We presuppose this hypothesis also along \emph{global paths of light} which can be thought of as a connected covering of multiple local segments.

\subsection{Light clock}\label{Kap - SRT Massbestimmung - light clock}

\begin{figure}         %Bem.: height skaliert Dateibild auf Zielgroesse im Dokument
  \begin{center}         %eps Dateien sind einfuegbar und auch in dvi sichtbar
                         %pdf nur bei direkter pdf-Kompelierung sichtbar
  \includegraphics[height=9.3cm]{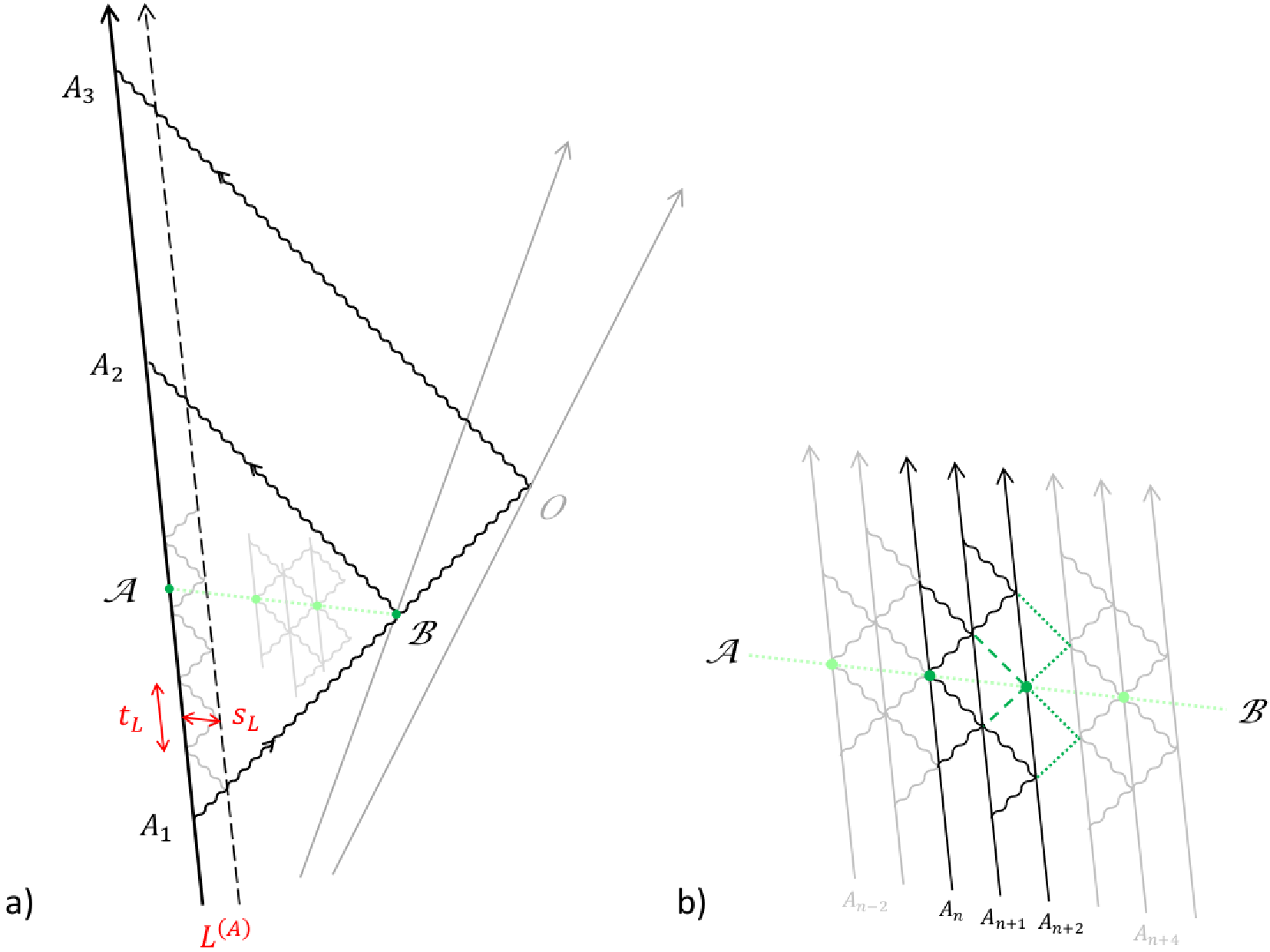}  %\\ if second picture aligned %below...
  \end{center}
  \vspace{-0.0cm}
  \caption{\label{figure-2} a) laser ranging b) consecutive and adjacent connection of light clocks
    }
  \end{figure}

Because of the universal light principle ''laser ranging'' is a reliable \emph{practice of navigation}.\footnote{The procedure developed naturally. Throughout millennia of evolution bats, coordinating their living at night, or dolphins, hunting under invisible conditions, discovered and rehearsed the practice of (i) \emph{producing} sonar waves and (ii) exploiting that \emph{tool} to master given living conditions.

Upon developing the classical metric one understands why it works so reliably in practice. With rulers and clocks one can measure the prerequisites. For durations of each sonar ranging act the emitting organism represents a sufficiently rigid body at constant motion. Sound propagates much faster, sufficiently straight and uniform. Thus by successive echoing animals can maneuver within an environment of comparably small relative motions. Based on common navigation actions Einstein demonstrates standardization of the conduct of spatiotemporal measurements. He discovered the Light principle as extra condition for physical measurements. Its theoretical conception led engineers into a revolution of technical applications (GPS, Lunar-Laser-Ranging, synchronization and coordination of partition of work on a global scale etc.).}
Let $\mathcal{A}$lice send out light towards $\mathcal{O}$tto\; $\mathcal{A}_1\!\rightsquigarrow\mathcal{O}\rightsquigarrow\mathcal{A}_2$ and towards $\mathcal{B}$ob\; $\mathcal{A}_1\!\rightsquigarrow\mathcal{B}\rightsquigarrow\mathcal{A}_3$ and wait until their reflection returns (see figure
%consecutive sequence of light clocks
\ref{figure-2}a). In radar round trips we focus on the distance covered and $\mathcal{A}$lice waiting time. For two ranging cycles $\mathcal{A}_1\!\rightsquigarrow\mathcal{O}\rightsquigarrow\mathcal{A}_2$ and $\mathcal{A}_1\!\rightsquigarrow\mathcal{B}\rightsquigarrow\mathcal{A}_3$ $\mathcal{A}$lice notices the order in which the light returns. By the light principle more waiting time $t_{\overline{\mathcal{A}_1\mathcal{A}_2}}>t_{\overline{\mathcal{A}_1\mathcal{A}_3}}$ corresponds to a larger distance covered $s_{\overline{\mathcal{AO}}}>s_{\overline{\mathcal{AB}}}$ from $\mathcal{A}$lice to turning point $\mathcal{O}$tto resp. $\mathcal{B}$ob and back.

For quantification $\mathcal{A}$lice constructs a reference device,
%[measurement unit--zur reproduzierbarkeit des Vergleiches waehle Standard/unit]
a \emph{light clock} $\mathcal{L} : \mathcal{L}_I\rightsquigarrow \mathcal{L}_{II} \rightsquigarrow \mathcal{L}_I \ldots$ with two nearby mirrors $\mathcal{L}_{I}$ and $\mathcal{L}_{II}$ in a rigid frame. The light constantly oscillates back and forth. Each tick of her \emph{measurement unit} $\mathcal{L}$ covers the same \emph{standard distance} $s_{\mathcal{L}}$ and takes the same \emph{standard time} $t_{\mathcal{L}}$ \{\ref{Kap - SRT Massbestimmung - light principle}\}.

$\mathcal{A}$lice light clock substitutes the traditional rulers and clocks. The protophysical test norms - for manufacturing both, traditional clocks and light clocks - remain the same. In practice one replaces the former by new light clocks because they realize the aspired ideal of uniform running more precisely. The motion of light is not anymore measured with traditional rulers and clocks; instead we determine all other motions with respect to the motion of light in (classical protophysically manufactured) light clocks. The light principle implies a \emph{paradigm shift}. One abandons the former priority of classical measurement devices in favor for the universal propagation of light. The motion of light becomes a measurement standard itself.\footnote{The definition of standard length $s_{\mathcal{L}}$ is based on a \emph{given} standard duration $t_{\mathcal{L}}$ and the universal speed of light $c$. Contemporary metrology regards speed of light $c$ as invariant natural constant and introduces optical clocks as frequency standards. Our world's current time standard (a laser-cooled cesium fountain known as NIST-F1 based on resonant transitions between quantized energy levels in atoms) is accurate to within $\Delta f/f \sim 10^{-16}$. It represents the ultimate reference for time intervals $t_{\mathrm{Cs}}$ with accuracy $~10^{-16}\mathrm{sec}$.

Bureau of Standards defines the atomic second $1\mathrm{sec}^{(\mathrm{SI})}:= 9192631770 \cdot t_{\mathrm{Cs}}$ - on paper - as a multiple of that standard duration. ''Cesium provides a 'physical' second that can be realized in laboratories and used for other measurements. ... The basic principle of the atomic oscillator is simple: Since all atoms of a specific element are identical, they should produce the exact same frequency...'' \cite{NIST - Frequency Standards and Realization of SI Second}. They refer to an \emph{intrinsic} property of an atom under standardized conditions: ''cesium atom at rest at a thermodynamic temperature of $0\mathrm{K}$''. An unperturbed atomic transition is identical from atom to atom (reproducibility).} We use the classical light clock as a new \emph{measurement unit}.

\subsection{Direct connections}\label{Kap - SRT Massbestimmung - direct connecting products}

We essentially refer to the oscillating light inside. In practice the two dimensions ''length'' and ''duration'' of a light clock $\mathcal{L}$ are always addressed unified. Depending on the concatenation:
\begin{itemize}
   \item   adjacently connected $\ast_s$ (ticking) light clocks represent a distance unit $\mathcal{L}_s$ and
   \item   consecutively connected $\ast_t$ (light clock) ticks represent a duration unit $\mathcal{L}_t$.
\end{itemize}
The width of the light clock $\mathcal{L}$ becomes our \emph{unit length} $s_{\mathcal{L}}$ and each tick lasts \emph{unit time} $t_{\mathcal{L}}$.

%By matching the independent light rays between adjacent light clocks $\mathcal{A}$lice \emph{connects} a swarm of comoving light clocks. From solely congruent building blocks she assembles a grid which covers larger objects or processes (e.g. $\mathcal{B}$ob's relative motion). By counting them she determines the magnitude of distances or durations. $\mathcal{A}$lice measures the relative motion of $\mathcal{B}$ob by the number and layout of units $\mathcal{L}$ and their reference length resp. duration.

\subsubsection{Time-like concatenation}

Let $\mathcal{A}$lice join together light clock ticks $\mathcal{L}$ \emph{one after another} until the sequence - symbolized by $\mathcal{L}\ast_{t}\ldots\ast_{t}\mathcal{L}$ - covers the waiting interval of her laser ranging cycle $\mathcal{A}_1\!\rightsquigarrow\mathcal{B}\rightsquigarrow\mathcal{A}_2$
\be\label{Formel - radar duration konstruierbare Ersetzung}
   \overline{\mathcal{A}_1\mathcal{A}_2} \; \sim_{t} \;
   \mathcal{L}\ast_{t}\ldots\ast_{t}\mathcal{L}  \;\; .
\ee
In her material model $\mathcal{A}$lice can count the number of ticks, symbolized $\sharp \left\{ \mathcal{L}_t \right\} =: t^{(\mathcal{A})}_{\overline{\mathcal{A}_1\mathcal{A}_2}}\,$. The ordering relation ''longer than'' becomes quantified. $\mathcal{A}$lice measures the duration of her laser ranging interval
\be\label{Formel - radar duration physical measure}
   t_{\overline{\mathcal{A}_1\mathcal{A}_2}} \;\;
   \stackrel{(\ref{Formel - radar duration konstruierbare Ersetzung})}{=} \;\;
   t_{\mathcal{L}\ast_t\ldots\ast_t\mathcal{L}} \;\; \stackrel{(\mathrm{Congr.})}{=:} \;\;
   t^{(\mathcal{A})}_{\overline{\mathcal{A}_1\mathcal{A}_2}} \; \cdot \; t_{\mathcal{L}}
\ee
by the sequence of ticks and the latter according to the congruence principle by the number of congruent (light clock) ticks and its reference duration  $t_{\mathcal{L}}$.

\subsubsection{Space-like concatenation}

Furthermore $\mathcal{A}$lice can place ticking light clocks $\mathcal{L}$ literally \emph{side by side}. She utilizes the same \emph{units} $\mathcal{L}$ to produce an adjacent layout of comoving light clocks.
\begin{lem}\label{Lem - SRT Kin - construct straight measurement path}
It represents $\mathcal{A}$lice \underline{simultaneous straight measurement path} towards $\mathcal{B}$ob $\overline{\mathcal{AB}}$.
\end{lem}
\textbf{Proof:}
Imagine a swarm of identically constituted light clocks $\mathcal{L}^{(\mathcal{A}_i)}$. Beginning with her own in moment $\mathcal{A}$ $\mathcal{A}$lice successively places pairs of light clocks $\mathcal{L}^{(\mathcal{A}_i)} \ast_s \mathcal{L}^{(\mathcal{A}_{i+1})}$ next to one another by letting their inner light rays overlap. She builds a locally regular grid of light clocks in an intrinsically \emph{simultaneous and straight way} (see figure
%adjacent layout of light clocks
\ref{figure-2}b):
\begin{enumerate}
\item[(a)]   Suppose we have successively laid out light clocks from $\mathcal{L}^{(\mathcal{A}_1)}$ all the way to $\mathcal{L}^{(\mathcal{A}_n)}$. Consider the two ticks of light clock $\mathcal{L}^{(\mathcal{A}_n)} \: \ast\!\mid_{\mathcal{A}_n}\mathcal{L}^{(\mathcal{A}_n)}$ around the moment $\mathcal{A}_n$.
\item[(b)]   The next comoving light clock $\mathcal{L}^{(\mathcal{A}_{n+1})}$ is placed so that the extended (dashed) light ray from $\mathcal{L}^{(\mathcal{A}_n)} \: \ast\!\mid_{\mathcal{A}_n}\mathcal{L}^{(\mathcal{A}_n)}$ coincides with the light ray from $\mathcal{L}^{(\mathcal{A}_{n+1})}$.
\item[(c)]   Starting from $\mathcal{A}_{n+1}$ - by isotropy - light travels in identical \emph{round trip duration} $t_{\mathcal{L}^{(\mathcal{A}_{n+1})}}$ the same distance back to left light clock $\mathcal{L}^{(\mathcal{A}_{n})}$ as to the right light clock $\mathcal{L}^{(\mathcal{A}_{n+1})}$.\footnote{Let two synchronously ticking light clocks $\mathcal{L} \ast_s \mathcal{L}$ sit side by side. We assume that \emph{two-way} light cycle on the left covers same standard distance $s_{\mathcal{L}}$ to its turning point as the other \emph{two-way} light cycle on the right.}
\item[(d)]   Consider a series of three (preceding and following) ticks of light clock $\mathcal{L}^{(\mathcal{A}_{n+1})}$.
\item[(e)]   Place light clock $\mathcal{L}^{(\mathcal{A}_{n+2})}$ so that the extended (dotted) light rays from $\mathcal{L}^{(\mathcal{A}_{n+1})}$ coincide with the two ticks of light clock $\mathcal{L}^{(\mathcal{A}_{n+2})} \: \ast\!\mid_{\mathcal{A}_{n+2}}\mathcal{L}^{(\mathcal{A}_{n+2})}$ around the moment $\mathcal{A}_{n+2}$.
\item[(f)]   By analogous induction steps $\;\mathcal{L}^{(\mathcal{A}_n)} \ast\!\mid_{\mathcal{A}_n}\mathcal{L}^{(\mathcal{A}_n)} \: \Rightarrow \: \mathcal{L}^{(\mathcal{A}_{n+1})} \: \Rightarrow \: \mathcal{L}^{(\mathcal{A}_{n+2})} \: \ast\!\mid_{\mathcal{A}_{n+2}}\mathcal{L}^{(\mathcal{A}_{n+2})}  \; \forall n$
    $\mathcal{A}$lice proceeds towards $\mathcal{B}$ob.

    In every step the extended (dashed resp. dotted) light rays coincide. In her straight comoving \emph{connection} $\mathcal{L}^{(\mathcal{A})}\!\!\mid_{\mathcal{A}} \ast \; \mathcal{L}^{(\mathcal{A}_2)} \ast \; \mathcal{L}^{(\mathcal{A}_3)}\!\!\mid_{\mathcal{A}_3}  \ldots \mathcal{L}^{(\mathcal{A}_n)}\!\!\mid_{\mathcal{A}_n} \ast \; \mathcal{L}^{(\mathcal{A}_{n+1})} \ast \; \mathcal{L}^{(\mathcal{A}_{n+2})}\!\!\mid_{\mathcal{A}_{n+2}}$ all light clocks tick synchronized
    along connecting moments $\mathcal{A}\:, \mathcal{A}_3 \ldots \mathcal{A}_n, \mathcal{A}_{n+2} \ldots \mathcal{B}$.
\end{enumerate}
The construction steps do not depend on the scale of light clock $\mathcal{L}$ (e.g. refining the locally regular layout with twice the light clocks of half the size coincides with the original pattern). They are locally well-defined; the global measurement path $\mathcal{L}\ast_s\ldots\ast_s\mathcal{L}$ is universal.
\qed
$\mathcal{A}$lice covers the laser ranging path to $\mathcal{B}$ob by an adjacent layout of light clocks
\be\label{Formel - radar distance konstruierbare Ersetzung}
   \overline{\mathcal{A}\mathcal{B}} \; \sim_{s} \;   \mathcal{L}\ast_s\ldots\ast_s\mathcal{L} \;\; .
\ee
Each represents a length unit $\mathcal{L}_s$. $\mathcal{A}$lice measures the length along her laser ranging path
\be\label{Formel - radar distance - direct physical measure}
   s_{\overline{\mathcal{AB}}} \;\;
   \stackrel{(\ref{Formel - radar distance konstruierbare Ersetzung})}{=} \;\;
   s_{\mathcal{L}\ast_s\ldots\ast_s\mathcal{L}}
   \;\; \stackrel{(\mathrm{Congr.})}{=:} \;\; s^{(\mathcal{A})}_{\overline{\mathcal{AB}}} \cdot s_{\mathcal{L}}
\ee
by the adjacent layout  $\mathcal{L}\ast_s\ldots\ast_s\mathcal{L}$  and the latter according to the congruence principle by the number $\sharp \left\{ \mathcal{L}_s \right\} =:  s^{(\mathcal{A})}_{\overline{\mathcal{AB}}}\,$ of congruent clocks $\mathcal{L}_s$ and its standard length $s_{\mathcal{L}}$.

\subsubsection{Spacetime-like concatenation}

With every laser ranging ping $\mathcal{A}_1\!\rightsquigarrow\mathcal{B}\rightsquigarrow\mathcal{A}_2$ $\mathcal{A}$lice measures the position of $\mathcal{B}$ob at the moment $\mathcal{B}$ when her signal reflects (see figure \ref{figure-2}a). $\mathcal{A}$lice covers the outgoing light ray $\overline{\mathcal{A}_1\mathcal{B}}$ by a swarm of light clocks in both space-like and time-like way: She connects a consecutive sequence until ''half-time'' moment $\mathcal{A}$ (after waiting half of her laser ranging interval)
\[
   \overline{\mathcal{A}_1\mathcal{A}} \; \sim_t \;
   \mathcal{L}\!\!\mid_{\mathcal{A}_1} \ast_t \ldots \ast_t \mathcal{L}\!\!\mid_{\mathcal{A}}
\]
in light clock $\mathcal{L}\!\!\mid_{\mathcal{A}}$ to an adjacent layout of (ticking) light clocks that reaches to moment $\mathcal{B}$
\[
   \overline{\mathcal{A}\mathcal{B}} \; \sim_s \;
   \mathcal{L}\!\!\mid_{\mathcal{A}} \ast_s \ldots \ast_s \mathcal{L}\!\!\mid_{\mathcal{B}} \;\; .
\]
The collective motion of light inside the composite of ticking light clocks - symbolized by $\mathcal{L}\!\!\mid_{\mathcal{A}_1} \ast_t \ldots \ast_t \mathcal{L}\!\!\mid_{\mathcal{A}} \ast_s \ldots \ast_s \mathcal{L}\!\!\mid_{\mathcal{B}}$ - covers the light ray from $\mathcal{A}$lice towards $\mathcal{B}$ob
\be\label{Formel - radar spatiotemporal distance konstruierbare Ersetzung}
   \overline{\mathcal{A}_1\mathcal{B}} \; \sim_{t,s} \; \mathcal{L} \ast_t \ldots \ast_t \mathcal{L} \ast_s \ldots \ast_s \mathcal{L} \;\; .
\ee
$\mathcal{A}$lice utilizes copies of the same light clock $\mathcal{L}$ as spatiotemporal units. Along a consecutive segment $\mathcal{L} \ast_t \ldots \ast_t \mathcal{L}$ each represents a time unit $\mathcal{L}_t$ and along an adjacent segment $\mathcal{L} \ast_s \ldots \ast_s \mathcal{L}$ a distance unit $\mathcal{L}_s$. In both segments she counts the congruent ticks $\sharp \left\{ \mathcal{L}_t \right\} =: t^{(\mathcal{A})}_{\overline{\mathcal{A}_1\mathcal{B}}}$ and the congruent clocks $\sharp \left\{ \mathcal{L}_s \right\} =: s^{(\mathcal{A})}_{\overline{\mathcal{A}_1\mathcal{B}}}\,$. $\mathcal{A}$lice measures the spatiotemporal distance towards $\mathcal{B}$ob
\be \label{Formel - radar direkt spatiotemporal physical measure}
   (t,s)_{\overline{\mathcal{A}_1\mathcal{B}}}
   \;\; \stackrel{(\ref{Formel - radar spatiotemporal distance konstruierbare Ersetzung})}{=} \;\;
   (t,s)_{\mathcal{L} \ast_t \ldots \ast_t \mathcal{L} \ast_s \ldots \ast_s \mathcal{L}}
   \;\; \stackrel{(\ref{Formel - radar duration physical measure})(\ref{Formel - radar distance - direct physical measure})}{=} \;\; \left(\; t^{(\mathcal{A})}_{\overline{\mathcal{A}_1\mathcal{A}}} \cdot t_{\mathcal{L}} \; , \; s^{(\mathcal{A})}_{\overline{\mathcal{A}\mathcal{B}}} \cdot s_{\mathcal{L}} \; \right)
\ee
by her composite layout. It is reproducible from the number of congruent light clocks, the consecutive or adjacent way of their connection and their standard length $s_{\mathcal{L}}$ and duration $t_{\mathcal{L}}$.
\\

\begin{figure}         %Bem.: height skaliert Dateibild auf Zielgroesse im Dokument
  \begin{center}         %eps Dateien sind einfuegbar und auch in dvi sichtbar
                         %pdf nur bei direkter pdf-Kompelierung sichtbar
  \includegraphics[scale=0.48]{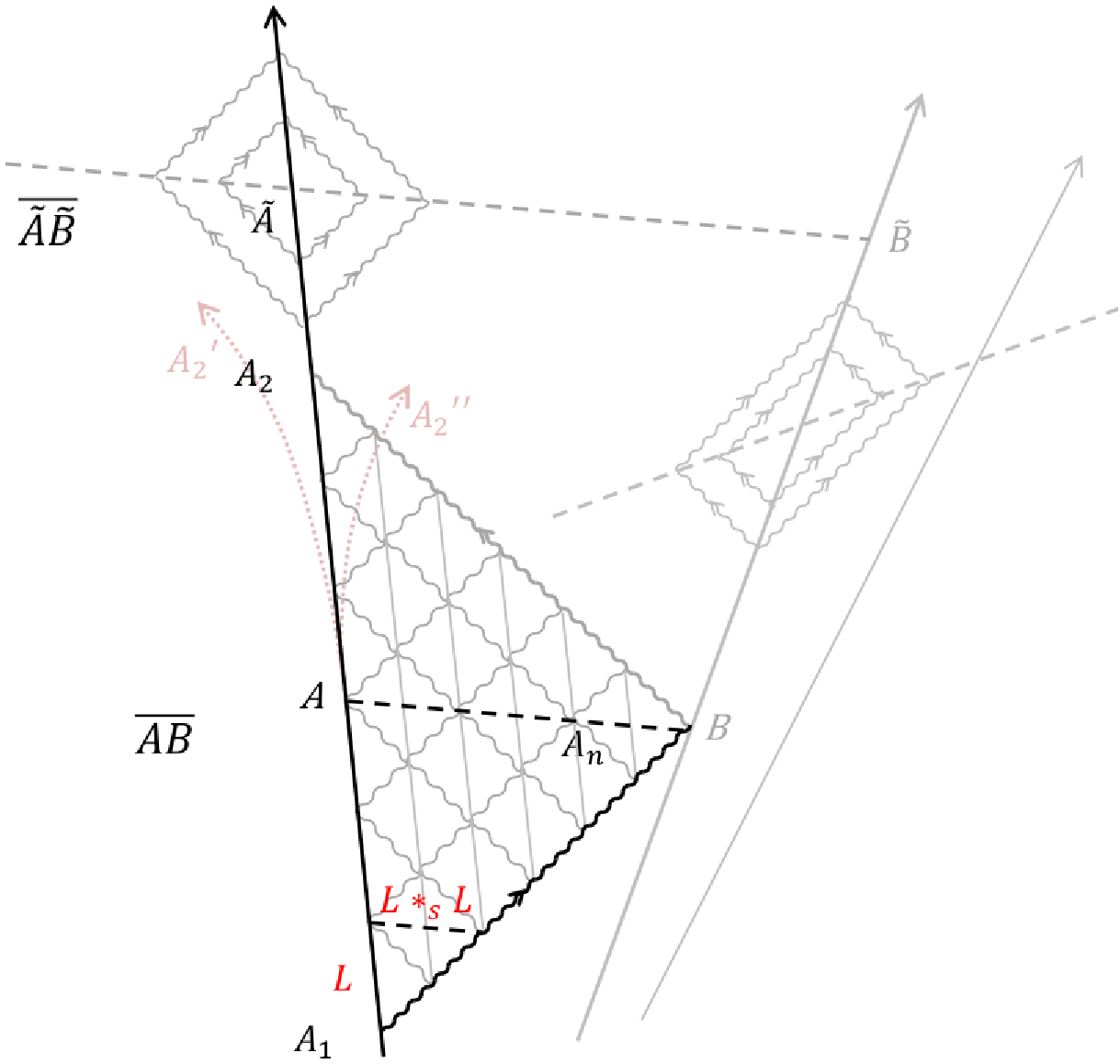}  %\\ if second picture aligned %below...
  \end{center}
  \vspace{-0.0cm}
  \caption{\label{figure-3} (local) indirect characterization of simultaneous straight measurement paths
    }
  \end{figure}

In the \emph{direct} measurement we cover the object or process by a grid of light clocks. Now consider the measurement of a ray of light. In figure \ref{figure-3}
$\mathcal{A}$lice covers the smallest segment $\overline{\mathcal{A}_1\mathcal{B}'} \sim_{t,s} \mathcal{L}_t \ast (\mathcal{L} \ast_s \mathcal{L})$ with one light clock tick and two adjacent light clocks. Step by step she covers the uniform motion of the outgoing light ray $\overline{\mathcal{A}_1\mathcal{B}}$ by a (locally) regular pattern of light clocks (and the same for the returning light ray $\overline{\mathcal{B}\mathcal{A}_2}$). No matter to what extent $\mathcal{A}$lice covers the departing light ray $\overline{\mathcal{A}_1\mathcal{B}'}\subset\overline{\mathcal{A}_1\mathcal{B}}$ by similarity any pair of durations scales proportional with the corresponding pair of distances\footnote{The division in $s_1/s_2$ resp. $t_1/t_2$ symbolizes $\mathcal{A}$lice dividing \emph{operation}. The formulation $s_{\overline{\mathcal{A}\mathcal{B}}}/s_{\mathcal{L}}:=n$ means: by connecting $n$ congruent reference paths $\mathcal{L} \ast_s \ldots \ast_s \mathcal{L} \sim_{s} \overline{\mathcal{A}\mathcal{B}}$ she will cover the path $\overline{\mathcal{A}\mathcal{B}}$.}
\be \label{Formel - radar direkt measurement c - proportionality}
   \frac{s_{\overline{\mathcal{A}_1\mathcal{B}}}}{s_{\overline{\mathcal{A}_1\mathcal{B}'}}} \;\; = \;\; \frac{t_{\overline{\mathcal{A}_1\mathcal{B}}}}{t_{\overline{\mathcal{A}_1\mathcal{B}'}}}  \;\; .
\ee
For a generic segment $(t_c,s_c)$ of the uniform light ray we express the proportionality relation $\frac{s_c}{2s_{\mathcal{L}}} = \frac{t_c}{t_{\mathcal{L}}}$ between pairs of \emph{same} basic observables by a proportionality constant
\be \label{Formel - radar direkt measurement c - proportionality constant}
   \left\{ \frac{s_c}{s_{\mathcal{L}}} \right\} \;\; \stackrel{(\ref{Formel - radar direkt measurement c - proportionality})}{=} \;\; \underbrace{2}_{\equiv \: c^{(\mathcal{L})}} \; \cdot \; \left\{ \frac{t_c}{t_{\mathcal{L}}} \right\}   \;\; .
\ee
We define the \emph{velocity} of light $c^{(\mathcal{L})} := s_c^{(\mathcal{L})}\! / t_c^{(\mathcal{L})} = 2$ (in standard light clock dimensions $s_{\mathcal{L}}, t_{\mathcal{L}}$) as a derived physical quantity. From known ''distance \emph{for each} time \emph{unit}'' $2\cdot s_{\mathcal{L}}$ and the ''number of time units'' $\frac{t_c}{t_{\mathcal{L}}}$ along the way one gets the total distance as a product of velocity\footnote{The formal expression $\frac{s_{\mathcal{L}}}{t_{\mathcal{L}}}$ has no \emph{physical} meaning. One cannot divide a path by a time \cite{Wallot - Groessengleichungen Einheiten und Dimensionen}. The formal reduction of fractions, that ''same dimensions (unit length, unit mass etc.) cancel one another'', gives back the relation (\ref{Formel - radar direkt measurement c - proportionality constant}) between quantities (ratios) which can all be measured directly by concatenation operations.} and time of flight
\be\label{Formel - radar direkt measurement c}
   s_c \;\; \stackrel{(\ref{Formel - radar direkt measurement c - proportionality constant})}{=} \;\; 2\cdot s_{\mathcal{L}} \cdot \frac{t_c}{t_{\mathcal{L}}}
   \;\; =: \;\; \underbrace{\left( 2\cdot \frac{s_{\mathcal{L}}}{t_{\mathcal{L}}} \right)}_{\equiv \: c} \;\;\cdot\;\; t_c  \;\; .\footnote{For measuring $c = c^{(\mathcal{L})} \cdot \frac{s_{\mathcal{L}}}{t_{\mathcal{L}}}$ we utilize a light clock with dimensions: width $s_{\mathcal{L}}$ and cycle length $t_{\mathcal{L}}$. Basic dimensions (unit length, unit time etc.) are ''arbitrarily chosen constant reference measures'' \cite{Wallot - Groessengleichungen Einheiten und Dimensionen}. Contemporary metrology refers to units based on the standard duration $t_{\mathrm{Cs}}$ of an intrinsic Cesium period and the invariant speed of light $c$ \{\ref{Kap - SRT Massbestimmung - light clock}\}. The atomic second $\mathrm{sec}_{\mathrm{SI}} := 9192631770 \cdot t_{\mathrm{Cs}}$ is a multiple of that standard duration (factor chosen to match traditional calendrical second). The (multiple of the) standard meter $299792458\cdot \mathrm{m}_{\mathrm{SI}} := (c \cdot  \mathrm{sec}_{\mathrm{SI}})$ is the distance of free light in one atomic second of flight. The numerical factor is fixed by convention (to cover $1 \mathrm{m}_{\mathrm{SI}} \simeq 1 \mathrm{m}_{\mathrm{bar}}$ the traditional platinum-iridium standard in the Bureau of Weights and Measures). One refers to the traditional units (meter bar and fraction of a tropical year) one last time, to match the conversion factors. Now one defines the international unit measures $\mathrm{sec}_{\mathrm{SI}}$, $\mathrm{m}_{\mathrm{SI}}$ independently and more precise from invariant natural processes (intrinsic $\mathrm{Cs}$-period and speed of light) and the fixed numerical factors; the old prototypes stay in the museum.

A light clock with SI-unit period $t_{\mathcal{L}}:=t_{\mathrm{SI}}$ and corresponding width $2 \cdot s_{\mathcal{L}} = (c \cdot 1 \mathrm{sec}_{\mathrm{SI}}) =299792458\cdot \mathrm{m}_{\mathrm{SI}}$ has the proportionality constant $\left\{ \frac{s_c}{\mathrm{m}} \right\}  \stackrel{(\ref{Formel - radar direkt measurement c - proportionality constant})}{=}  299792458 \; \cdot \; \left\{ \frac{t_c}{\mathrm{sec}} \right\}$; thus measures speed of light $c = 299792458 \cdot \frac{\mathrm{m}}{\mathrm{sec}}$.}
\ee

\subsection{Indirect laser ranging}\label{Kap - SRT Massbestimmung - indirect laser ranging}

In direct laser ranging $\mathcal{A}_1\!\rightsquigarrow\mathcal{B}\rightsquigarrow\mathcal{A}_2$ $\mathcal{A}$lice measures the distance $s_{\overline{\mathcal{AB}}}$ to $\mathcal{B}$ob (\ref{Formel - radar distance - direct physical measure}) by counting units along the (potentially global) simultaneous measurement path $\overline{\mathcal{AB}}$. From a direct measurement the departing and returning ray of light\footnote{Locally regular pattern of light clocks (see figure \ref{figure-3}) covers laser ranging waiting interval $\overline{\mathcal{A}_1\mathcal{A}_2}$ by same number of consecutive (light clock) ticks as there are adjacent (ticking) clocks along laser ranging route $\overline{\mathcal{AB}}$.} cover in the same duration $t_{\overline{\mathcal{A}_1\mathcal{B}}} = t_{\overline{\mathcal{B}\mathcal{A}_2}} = \frac{1}{2} \cdot t_{\overline{\mathcal{A}_1\mathcal{A}_2}}$ the proportional distance $s_{\overline{\mathcal{A}\mathcal{B}}} \stackrel{(\ref{Formel - radar direkt measurement c})}{=} c \cdot t_{\overline{\mathcal{A}_1\mathcal{B}}} = \frac{c}{2} \cdot t_{\overline{\mathcal{A}_1\mathcal{A}_2}}$. Thus (locally) $\mathcal{A}$lice can also indirectly compute that length
\be \label{Formel - radar indirekt spatiotemporal physical measure}
   (t,s)_{\overline{\mathcal{A}_1\mathcal{B}}} \;\; \stackrel{(\ref{Formel - radar direkt spatiotemporal physical measure})(\ref{Formel - radar direkt measurement c})}{=} \;\; \left( \;\; \frac{1}{2} \cdot t_{\overline{\mathcal{A}_1\mathcal{A}_2}} \;\; , \;\; \frac{c}{2} \cdot t_{\overline{\mathcal{A}_1\mathcal{A}_2}} \;\; \right)
\ee
from measuring the round trip time $t_{\overline{\mathcal{A}_1\mathcal{A}_2}}$; the familiar principle of indirect laser ranging.

For the resulting equation $\mathcal{A}$lice must obey a \emph{measurement condition} (underlying the direct measurement of light rays in figure \ref{figure-3}), that during the radar waiting interval $\overline{\mathcal{A}_1\mathcal{A}_2}$ her motion is preserved. After emitting the light pulse $\overline{\mathcal{A}_1\mathcal{B}}$ she neither accelerates away $\mathcal{A}_2'$ nor towards $\mathcal{A}_2''$ the returning light pulse $\overline{\mathcal{B}\mathcal{A}_1}$. In local laser ranging practice accelerations are negligible; for larger configurations the effects accumulate. Then $\mathcal{A}$lice can characterize all elements of her \emph{simultaneity line} $\mathcal{A}_n \in \overline{\mathcal{AB}}$ by moments $\mathcal{A'}, \mathcal{A''} \in \overline{\mathcal{A}_1\mathcal{A}_2}$ along the waiting interval (see figure \ref{figure-3}). In local laser ranging $\mathcal{A'}\!\rightsquigarrow\mathcal{A}_n\rightsquigarrow\mathcal{A''}$ the preceding emission and subsequent reception are symmetric $t_{\overline{\mathcal{A'A}}} = t_{\overline{\mathcal{AA''}}}$ \emph{with respect to $\mathcal{A}$lice moment $\mathcal{A}$}.

\begin{figure}         %Bem.: height skaliert Dateibild auf Zielgroesse im Dokument
  \begin{center}         %eps Dateien sind einfuegbar und auch in dvi sichtbar
                         %pdf nur bei direkter pdf-Kompelierung sichtbar
  \includegraphics[scale=0.30]{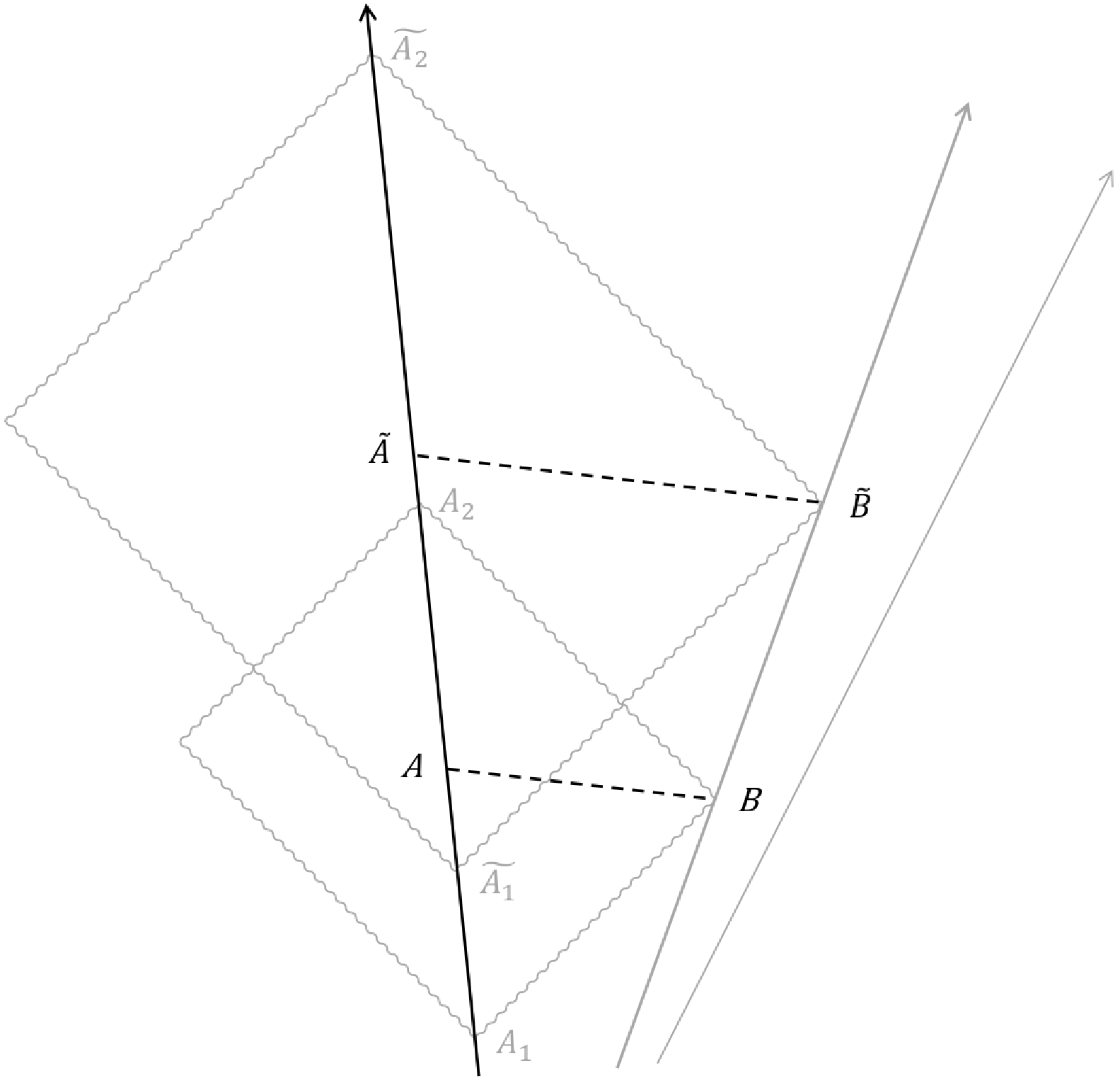}  %\\ if second picture aligned %below...
  \end{center}
  \vspace{-0.0cm}
  \caption{\label{figure-4} combination of two elementary laser ranging measurements
    }
  \end{figure}
With two elementary laser rangings $\mathcal{A}_1\!\rightsquigarrow\mathcal{B}\rightsquigarrow\mathcal{A}_2$ and $\tilde{\mathcal{A}_1}\!\rightsquigarrow\tilde{\mathcal{B}}\rightsquigarrow\tilde{\mathcal{A}_2}$ towards the two consecutive moments $\mathcal{B}$ and $\tilde{\mathcal{B}}$ (see figure \ref{figure-4}) $\mathcal{A}$lice measures the relative motion of $\mathcal{B}$ob $\overline{\mathcal{B}\tilde{\mathcal{B}}}$.\footnote{In a direct measurement her layout of light clocks $\mathcal{L}\!\!\mid_{\mathcal{B}} \ast_s \ldots \ast_s \mathcal{L}\!\!\mid_{\mathcal{A}} \ast_t \ldots \ast_t  \mathcal{L}\!\!\mid_{\tilde{\mathcal{A}}} \ast_s \ldots \ast_s \mathcal{L}\!\!\mid_{\tilde{\mathcal{B}}} \;\;\equiv\;\;  \overline{\mathcal{B}\mathcal{A}} \ast \overline{\mathcal{A}\tilde{\mathcal{A}}} \ast \overline{\tilde{\mathcal{A}}\tilde{\mathcal{B}}} \;\; \sim_{t,s} \;\; \overline{\mathcal{B}\tilde{\mathcal{B}}}$ covers a segment of his motion.} Her indirect laser ranging involves three steps:
\begin{enumerate}
\item   \emph{construct} her straight simultaneous measurement paths towards $\mathcal{B}$ob
\item   \emph{enclose} the measurement object $\overline{\mathcal{B}\tilde{\mathcal{B}}}$ by her simultaneity lines $\overline{\mathcal{AB}}$ and $\overline{\mathcal{\tilde{A}\tilde{B}}}$
    \be
       \overline{\mathcal{B}\tilde{\mathcal{B}}} \; \sim_{t,s} \; \overline{\mathcal{B}\mathcal{A}} \; \ast \; \overline{\mathcal{A}\tilde{\mathcal{A}}} \; \ast \; \overline{\tilde{\mathcal{A}}\tilde{\mathcal{B}}} \label{Formel - radar direkt motion Eingrenzung}
    \ee
\item   \emph{project} between and along the simultaneity lines for temporal and spatial components
\bea
   (t,s)_{\overline{\mathcal{B}\tilde{\mathcal{B}}}} & \stackrel{(\ref{Formel - radar direkt motion Eingrenzung})}{=} & (t,s)_{\overline{\mathcal{B}\mathcal{A}} \; \ast \; \overline{\mathcal{A}\tilde{\mathcal{A}}} \; \ast \; \overline{\tilde{\mathcal{A}}\tilde{\mathcal{B}}}} \nn \\
    & \stackrel{(\ref{Formel - radar direkt spatiotemporal physical measure})}{=} & \!\! \underbrace{(t,s)_{\overline{\mathcal{B}\mathcal{A}}}}_{ \left( 0 \; , \; - s_{\overline{\mathcal{A}\mathcal{B}}}\right)} \; + \; \underbrace{(t,s)_{\overline{\mathcal{A}\tilde{\mathcal{A}}}}}_{ \left( t_{\overline{\mathcal{A}\tilde{\mathcal{A}}}}  \; , \; 0 \right)} \; + \; \underbrace{(t,s)_{\overline{\tilde{\mathcal{A}}\tilde{\mathcal{B}}}}}_{ \left( 0 \; , \; s_{\overline{\tilde{\mathcal{A}}\tilde{\mathcal{B}}}}\right)}
    \;\; = \; \left( \; t_{\overline{\mathcal{A}\tilde{\mathcal{A}}}} \;\: , \; s_{\overline{\tilde{\mathcal{A}}\tilde{\mathcal{B}}}} - s_{\overline{\mathcal{A}\mathcal{B}}}  \; \right) \label{Formel - radar direkt motion physical measure}
\eea
\end{enumerate}
The vectorial addition of components $(t,s)_{\overline{\mathcal{A}\tilde{\mathcal{A}}}} + (t,s)_{\overline{\tilde{\mathcal{A}}\tilde{\mathcal{B}}}} = (t,s)_{\overline{\mathcal{A}\tilde{\mathcal{B}}}}\,$ corresponds with direct measurement operations. In the underlying material model we concatenate a number of light clocks $\mathcal{L} \ast_t \ldots \ast_t \mathcal{L}$ and $\mathcal{L} \ast_s \ldots \ast_s \mathcal{L}$ to create a composite layout $\mathcal{L} \ast_t \ldots \ast_t \mathcal{L} \ast_s \ldots \ast_s \mathcal{L}$.

Let $\mathcal{A}$lice and $\mathcal{B}$ob coincide (without loss of generality) in the initial moment $\mathcal{P}$. Now the first laser ranging configuration becomes trivial and we are left with $\mathcal{P}\rightarrow \mathcal{A}_1\!\rightsquigarrow\mathcal{B}\rightsquigarrow\mathcal{A}_2$ (see figure \ref{figure-5}). $\mathcal{A}$lice measures the spatiotemporal interval of $\mathcal{B}$ob's \emph{motion}
\be
   (t,s)_{\overline{\mathcal{PB}}} \;\; \stackrel{(\ref{Formel - radar direkt motion physical measure})(\ref{Formel - radar indirekt spatiotemporal physical measure})}{=} \;\; \left( \; t_{\overline{\mathcal{P}\mathcal{A}_1}} \; + \; \frac{1}{2} \cdot t_{\overline{\mathcal{A}_1\mathcal{A}_2}}   \;\; , \;\;
      \frac{c}{2} \cdot t_{\overline{\mathcal{A}_1\mathcal{A}_2}}
   \; \right) \label{Formel - radar indirekt motion physical measure} \;\; .
\ee

%%%%%%%%%%%%%%%%%%%%%%%%%%%%%%%%%%%%%%%%%%%%%%%%%%%%%%%%%%%%%%%%%%%%%%%%%

\section{Lorentz transformation}\label{Kap - Masszsh}

\begin{figure}         %Bem.: height skaliert Dateibild auf Zielgroesse im Dokument
  \begin{center}         %eps Dateien sind einfuegbar und auch in dvi sichtbar
                         %pdf nur bei direkter pdf-Kompelierung sichtbar
  \includegraphics[height=19cm]{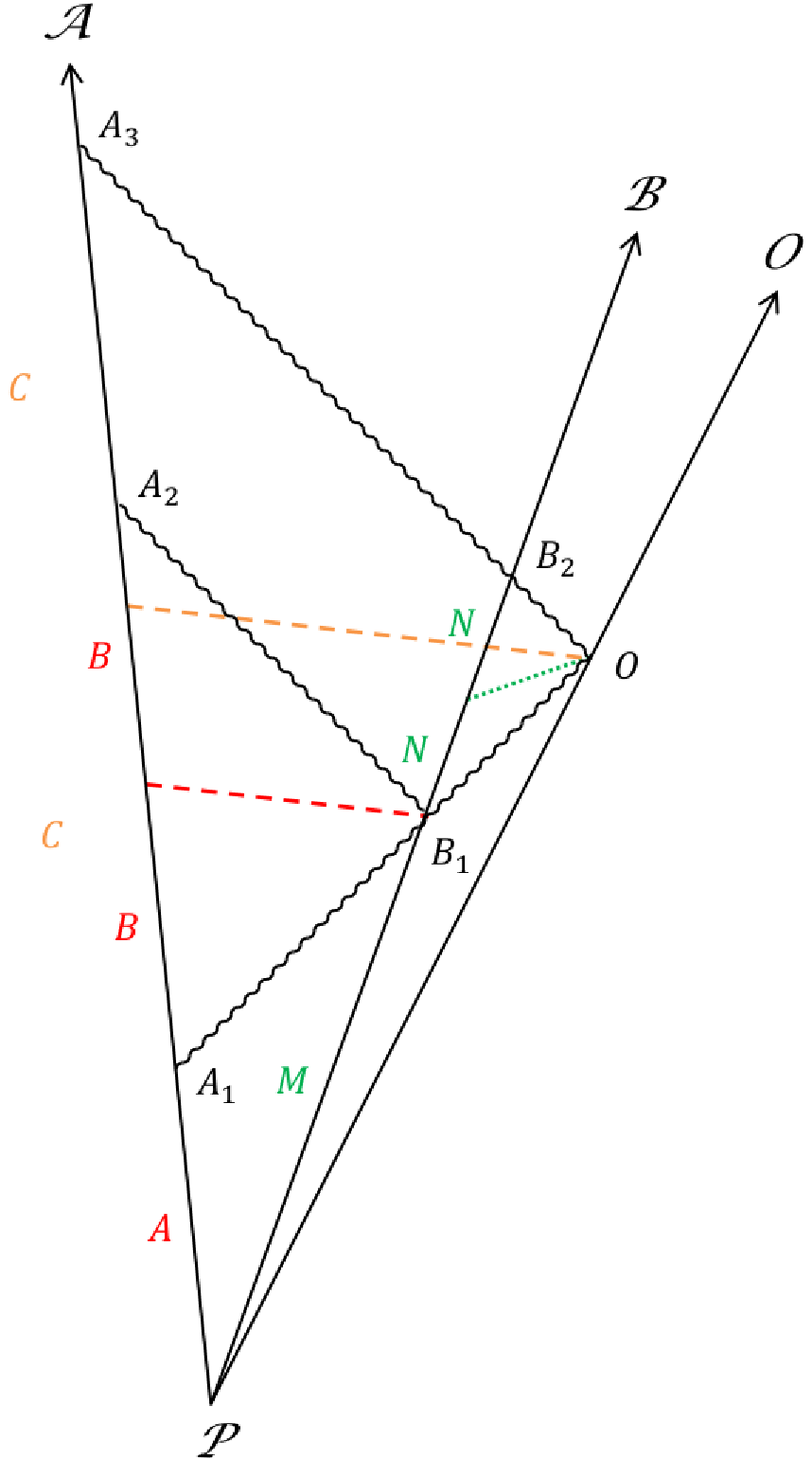}  %\\ if second picture aligned %below...
  \end{center}
  \vspace{-0.0cm}
  \caption{\label{figure-5} coinciding light rays in intrinsic measurements by $\mathcal{A}$lice and $\mathcal{B}$ob
    }
  \end{figure}

We have defined the termini of $\mathcal{A}$lice laser ranging measurements towards $\mathcal{O}$tto $\mathcal{P}\rightarrow\mathcal{A}_1\rightsquigarrow\mathcal{O}\rightsquigarrow\mathcal{A}_3$. Let another observer $\mathcal{B}$ob measure the same segment $\overline{\mathcal{PO}}$ of $\mathcal{O}$tto's motion (see figure \ref{figure-5}). $\mathcal{B}$ob conducts laser ranging $\mathcal{P}\rightarrow\mathcal{B}_1\rightsquigarrow\mathcal{O}\rightsquigarrow\mathcal{B}_2$ in the same way as $\mathcal{A}$lice. Following protophysical principles he manufactures his own light clock $\mathcal{L}^{(\mathcal{B})}$ and uses it in a standardized way. Step by step $\mathcal{B}$ob develops analogous measurement termini \{\ref{Kap - SRT Massbestimmung}\}.

$\mathcal{B}$ob \emph{constructs} his (dotted) simultaneity lines towards $\mathcal{O}$tto $\overline{\mathcal{BO}}$ (or back to $\mathcal{A}$lice $\overline{\mathcal{BA}}$). Directly, by adjacent connection of comoving light clocks $\mathcal{L}^{(\mathcal{B})} \ast_s \ldots \ast_s \mathcal{L}^{(\mathcal{B})}$, or indirectly, from round trip signaling times. Though, the \emph{same measurement principle} and intrinsic operations (independent propagation of light and intrinsic construction and connection of their respective light clocks) lead not to the same results. $\mathcal{A}$lice constructs simultaneity lines $\overline{\mathcal{AB}}$, $\overline{\mathcal{AO}}$ with different orientation than $\mathcal{B}$ob's simultaneity lines $\overline{\mathcal{BA}}$, $\overline{\mathcal{BO}}$ (see figure \ref{figure-3}).

Next $\mathcal{B}$ob \emph{encloses} measurement object $\mathcal{O}$tto $\overline{\mathcal{P}\mathcal{O}}$ in between his simultaneity lines $\overline{\mathcal{BO}}$
\[
   \overline{\mathcal{PO}} \;\; \stackrel{(\ref{Formel - radar direkt motion Eingrenzung})}{\sim_{t,s}}\;\; \overline{\mathcal{P}\mathcal{B}} \; \ast \; \overline{\mathcal{B}\mathcal{O}}
\]
and projects $\mathcal{O}$tto's relative motion onto the spatial and temporal components
\bea
   (t,s)_{\overline{\mathcal{PO}}} & = & (t,s)_{\overline{\mathcal{P}\mathcal{B}_1} \; \ast \; \overline{\mathcal{B}_1\mathcal{B}} \; \ast \; \overline{\mathcal{B}\mathcal{O}}} \nn \\
    & \stackrel{(\ref{Formel - radar direkt motion physical measure})}{=} & \left( \; t_{\overline{\mathcal{P}\mathcal{B}_1}} +  t_{\overline{\mathcal{B}_1\mathcal{B}}} \;\; , \;\; s_{\overline{\mathcal{B}\mathcal{O}}}  \; \right) \label{Formel - radar Bobs direkt motion physical measure of Otto} \;\;\; .
\eea

The intrinsic procedure is the same. By consecutive and adjacent connection of her light clocks $\mathcal{L}^{(\mathcal{A})}$ $\mathcal{A}$lice covers same segment of $\mathcal{O}$tto's motion as $\mathcal{B}$ob with his light clocks $\mathcal{L}^{(\mathcal{B})}$
\bea
   \overline{\mathcal{PO}} & \sim_{t,s} &  \;\; \left( \; t^{(\mathcal{A})}_{\overline{\mathcal{P}\mathcal{O}}} \;\: , \;\; s^{(\mathcal{A})}_{\overline{\mathcal{P}\mathcal{O}}} \; \right) \cdot \mathcal{L}^{(\mathcal{A})} \nn \\
    & \sim_{t,s} &  \;\; \left( \; t^{(\mathcal{B})}_{\overline{\mathcal{P}\mathcal{O}}} \;\: , \;\; s^{(\mathcal{B})}_{\overline{\mathcal{P}\mathcal{O}}} \; \right) \cdot \mathcal{L}^{(\mathcal{B})} \;\;\; . \nn
\eea
Let both also measure the same segment of $\mathcal{B}$ob's and of  $\mathcal{A}$lice' motion
\bea
   \overline{\mathcal{P}\mathcal{B}_1} & \sim_{t,s} &  \;\; \left( \; t^{(\mathcal{A})}_{\overline{\mathcal{P}\mathcal{B}_1}} \;\: , \;\; s^{(\mathcal{A})}_{\overline{\mathcal{P}\mathcal{B}_1}} \; \right) \cdot \mathcal{L}^{(\mathcal{A})} \nn \\
    & \sim_{t,s} &  \;\; \left( \; t^{(\mathcal{B})}_{\overline{\mathcal{P}\mathcal{B}_1}} \;\: , \;\; 0 \; \right) \cdot \mathcal{L}^{(\mathcal{B})}  \nn \\
    & & \nn \\
   \overline{\mathcal{P}\mathcal{A}_1} & \sim_{t,s} &  \;\; \left( \; t^{(\mathcal{A})}_{\overline{\mathcal{P}\mathcal{A}_1}} \;\: , \;\; 0 \; \right) \cdot \mathcal{L}^{(\mathcal{A})} \nn \\
    & \sim_{t,s} &  \;\; \left( \; t^{(\mathcal{B})}_{\overline{\mathcal{P}\mathcal{A}_1}} \;\: , \;\; s^{(\mathcal{B})}_{\overline{\mathcal{P}\mathcal{A}_1}} \; \right) \cdot \mathcal{L}^{(\mathcal{B})} \;\;\; . \nn
\eea
The coinciding light rays in their laser ranging processes are depicted in figure \ref{figure-5}.

From the interrelation of their physical prerequisites and the same measurement principle we derive the transformation $\left( \; t^{(\mathcal{A})}_{\overline{\mathcal{P}\mathcal{O}}} \; , \: s^{(\mathcal{A})}_{\overline{\mathcal{P}\mathcal{O}}} \; \right) \: \leftrightarrow \: \left( \; t^{(\mathcal{B})}_{\overline{\mathcal{P}\mathcal{O}}} \; , \: s^{(\mathcal{B})}_{\overline{\mathcal{P}\mathcal{O}}} \; \right)$ between $\mathcal{A}$lice and $\mathcal{B}$ob's measured values of the same measurement object: $\mathcal{O}$tto's motion $\overline{\mathcal{PO}}$. Provided measurements of their own motion $\overline{\mathcal{P}\mathcal{A}_1}$, $\overline{\mathcal{P}\mathcal{B}_1}$ it follows by successive substitution in three steps:
\[
\begin{array}{ccccc}
   \mathrm{\mathbf{I}} & \overbrace{\left( \; t^{(\mathcal{B})}_{\overline{\mathcal{P}\mathcal{O}}} \; , \; s^{(\mathcal{B})}_{\overline{\mathcal{P}\mathcal{O}}} \; \right)}^{\overline{\mathcal{PO}} \; \mathrm{measured \; by \;} \mathcal{B}\mathrm{ob}}
      &
    \left( \; M \; , \; N \; \right)  &  &
    \nn \\
   \mathrm{\mathbf{II}} &  & \left( \; M \; , \; N \; \right) & \left( \; A \; , \; B \; , \; C \; \right) &
    \nn \\
   \mathrm{\mathbf{III}} &  &  & \left( \; A \; , \; B \; , \; C \; \right) & \underbrace{\left(
    \left( \; t^{(\mathcal{A})}_{\overline{\mathcal{P}\mathcal{O}}} \; , \; s^{(\mathcal{A})}_{\overline{\mathcal{P}\mathcal{O}}} \; \right) \&
    \left( \; t^{(\mathcal{A})}_{\overline{\mathcal{P}\mathcal{B}_1}} \; , \; s^{(\mathcal{A})}_{\overline{\mathcal{P}\mathcal{B}_1}} \; \right) \right)}_{\overline{\mathcal{PO}} \; \& \; \overline{\mathcal{P}\mathcal{B}_1} \; \mathrm{measured \; by \;} \mathcal{A}\mathrm{lice}}
      \nn
\end{array}
\]
where we use the following abbreviations for indirect laser ranging measurements by $\mathcal{A}$lice
\[
   A := t^{(\mathcal{A})}_{\overline{\mathcal{P}\mathcal{A}_1}}  \;\;\;\;\;\;\;\;\;\;\;
   B := \frac{1}{2} \cdot t^{(\mathcal{A})}_{\overline{\mathcal{A}_1\mathcal{A}_2}} \;\;\;\;\;\;\;\;\;\;\;
   C := \frac{1}{2} \cdot t^{(\mathcal{A})}_{\overline{\mathcal{A}_1\mathcal{A}_3}}
\]
and by $\mathcal{B}$ob
\[
   M := t^{(\mathcal{B})}_{\overline{\mathcal{P}\mathcal{B}_1}}  \;\;\;\;\;\;\;\;\;\;\;
   N := \frac{1}{2} \cdot t^{(\mathcal{B})}_{\overline{\mathcal{B}_1\mathcal{B}_2}} \;\;\; .
\]

In \textbf{step I} we express $\mathcal{B}$ob's indirectly determined values of $\mathcal{O}$tto's motion $(t,s)^{(\mathcal{B})}_{\overline{\mathcal{PO}}}$ in terms of $\mathcal{B}$ob's direct measurements of round-trip signaling durations $M$, $N$
\bea
   t^{(\mathcal{B})}_{\overline{\mathcal{P}\mathcal{O}}}
   & \stackrel{(\ref{Formel - radar indirekt motion physical measure})}{=} & M + N \label{Formel - Masszsh - Substitution st MN i}\\
   s^{(\mathcal{B})}_{\overline{\mathcal{P}\mathcal{O}}}
   & \stackrel{(\ref{Formel - radar indirekt motion physical measure})}{=} & c \cdot N \label{Formel - Masszsh - Substitution st MN ii} \;\;\; .
\eea

In \textbf{step II} we express $\mathcal{B}$ob's measured durations $M$, $N$ in terms of $\mathcal{A}$lice' duration measurements $A$, $B$, $C$. In order to substitute the two physical quantities $M$, $N$ in terms of physical quantities $A$, $B$, $C$ we need two relations between corresponding measurements.

$\mathcal{A}$lice' and $\mathcal{B}$ob's laser ranging processes overlap in figure \ref{figure-5}. The outgoing light rays $\overline{\mathcal{A}_1\mathcal{B}_1}$ and $\overline{\mathcal{A}_1\mathcal{O}}$ partially coincide and the reflected light rays $\overline{\mathcal{B}_1\mathcal{A}_2}$ and $\overline{\mathcal{B}_2\mathcal{A}_3}$ are parallel \{\ref{Kap - SRT Massbestimmung - light principle}\}. From similar triangles $\mathcal{P}\mathcal{B}_1\mathcal{A}_2$ and $\mathcal{P}\mathcal{B}_2\mathcal{A}_3$ (all sides are pairwise parallel) we get one relation
\be\label{Formel - Masszsh - Substitution MN ABC i}
   \frac{M}{A+B+B} \; = \; \frac{M+N+N}{A+C+C} \;\;\; .
\ee

For a second relation we analyze the two triangles $\mathcal{P}\mathcal{A}_1\mathcal{B}_1$ and $\mathcal{P}\mathcal{B}_1\mathcal{A}_2$. We can regard each as ''degenerate trapezoid'', as a \emph{calibration} procedure by means of which $\mathcal{A}$lice and $\mathcal{B}$ob can compare their light clocks $\mathcal{L}^{(\mathcal{A})}$ and $\mathcal{L}^{(\mathcal{B})}$:
\begin{itemize}
\item   In $\mathcal{P}\mathcal{A}_1\mathcal{B}_1$  $\mathcal{A}$lice sends out two light signals along $\overline{\mathcal{P}\mathcal{A}_1} = A \cdot \mathcal{L}^{(\mathcal{A})}_t$ - the first at moment $P$ and the second  at moment $\mathcal{A}_1$ after $\sharp A$ ticks of her light clock - which $\mathcal{B}$ob receives along $\overline{\mathcal{P}\mathcal{B}_1} = M \cdot \mathcal{L}^{(\mathcal{B})}_t$ - at moments $P$ and $\mathcal{B}_1$ after $\sharp M$ ticks of his light clock.
\item   In $\mathcal{P}\mathcal{B}_1\mathcal{A}_2$  $\mathcal{B}$ob sends out two light signals along $\overline{\mathcal{P}\mathcal{B}_1} = M \cdot \mathcal{L}^{(\mathcal{B})}_t$ - at moments $P$ and $\mathcal{B}_1$ after $\sharp M$ ticks of his light clock - which $\mathcal{A}$lice receives along $\overline{\mathcal{P}\mathcal{A}_2} = (A+B+B) \cdot \mathcal{L}^{(\mathcal{A})}_t$ - the first in $P$ and the second in $\mathcal{A}_2$ after $\sharp (A+B+B)$ ticks of her light clock.
\end{itemize}
If $\mathcal{A}$lice and $\mathcal{B}$ob use identically constituted reference devices (light clocks made from same material) then both encounter the same dilation effect for each others relative motion. According to the \emph{relativity principle} both configurations are intrinsically similar. $\mathcal{A}$lice and $\mathcal{B}$ob have no way to specify absolute motion. By means of intrinsic measurements both determine the same ratio between the two durations for receiving both signals (heard from the other) and the duration of the sending interval (measured by themselves)
\be\label{Formel - Masszsh - Substitution MN ABC ii}
   \frac{M}{A} \; \stackrel{!}{=} \; \frac{A+B+B}{M} \;\;\; .
\ee

In \textbf{step III} we express $\mathcal{A}$lice laser ranging durations $A$, $B$, $C$ in terms of $\mathcal{A}$lice indirect determined values (\ref{Formel - radar indirekt motion physical measure}) for $\mathcal{O}$tto's motion $(t,s)^{(\mathcal{A})}_{\overline{\mathcal{PO}}}$ and for $\mathcal{B}$ob's motion $(t,s)^{(\mathcal{A})}_{\overline{\mathcal{PB}_1}}$
\bea
   A & = &  t^{(\mathcal{A})}_{\overline{\mathcal{P}\mathcal{O}}} \; - \; \frac{1}{c} \cdot s^{(\mathcal{A})}_{\overline{\mathcal{P}\mathcal{O}}} \label{Formel - Masszsh - Substitution ABC st i} \\
   & = &  t^{(\mathcal{A})}_{\overline{\mathcal{P}\mathcal{B}}} \; - \; \frac{1}{c} \cdot s^{(\mathcal{A})}_{\overline{\mathcal{P}\mathcal{B}}} \label{Formel - Masszsh - Substitution ABC st ii}\\
   B & = & \frac{1}{c} \cdot s^{(\mathcal{A})}_{\overline{\mathcal{P}\mathcal{B}}} \label{Formel - Masszsh - Substitution ABC st iii}\\
   C & = & \frac{1}{c} \cdot s^{(\mathcal{A})}_{\overline{\mathcal{P}\mathcal{O}}} \label{Formel - Masszsh - Substitution ABC st iv} \;\;\; .
\eea

After successive insertion of these three steps (see appendix A) we can express $\mathcal{B}$ob's physical quantities of $\mathcal{O}$tto's motion $(t,s)^{(\mathcal{B})}_{\overline{\mathcal{PO}}}$ in terms of $\mathcal{A}$lice measurements of $\mathcal{O}$tto's motion $(t,s)^{(\mathcal{A})}_{\overline{\mathcal{PO}}}$ and of the relative motion of $\mathcal{B}$ob $(t,s)^{(\mathcal{A})}_{\overline{\mathcal{PB}_1}}$
\bea\label{Formel - Masszsh - Lorentz Trafo st matrix}
   t^{(\mathcal{B})}_{\overline{\mathcal{P}\mathcal{O}}} & \stackrel{(\ref{Formel - Masszsh - Lorentz Trafo t})}{=} &
   \;\;\;\;\;\: \frac{1}{\sqrt{1-\frac{v_{\mathcal{B}}^2}{c^2}}} \cdot \;\; t^{(\mathcal{A})}_{\overline{\mathcal{P}\mathcal{O}}} \;\;\;\;\: - \;\; \frac{1}{\sqrt{1-\frac{v_{\mathcal{B}}^2}{c^2}}} \cdot \frac{v_{\mathcal{B}}}{c^2} \cdot \: s^{(\mathcal{A})}_{\overline{\mathcal{P}\mathcal{O}}} \\
   s^{(\mathcal{B})}_{\overline{\mathcal{P}\mathcal{O}}} & \stackrel{(\ref{Formel - Masszsh - Lorentz Trafo s})}{=} & \!\!
   - \; \frac{1}{\sqrt{1-\frac{v_{\mathcal{B}}^2}{c^2}}} \cdot v_{\mathcal{B}} \cdot \: t^{(\mathcal{A})}_{\overline{\mathcal{P}\mathcal{O}}}  \;\;\; + \;\;\;\;\;\; \frac{1}{\sqrt{1-\frac{v_{\mathcal{B}}^2}{c^2}}} \cdot \;\; s^{(\mathcal{A})}_{\overline{\mathcal{P}\mathcal{O}}}  \nn
\eea
where $\mathcal{A}$lice determines $\mathcal{B}$ob's relative velocity from $v_{\mathcal{B}} := s^{(\mathcal{A})}_{\overline{\mathcal{P}\mathcal{B}}} \!\left/\! t^{(\mathcal{A})}_{\overline{\mathcal{P}\mathcal{B}}}\right.$.

\begin{figure}[t!]         %Bem.: height skaliert Dateibild auf Zielgroesse im Dokument
  \begin{center}         %eps Dateien sind einfuegbar und auch in dvi sichtbar
                         %pdf nur bei direkter pdf-Kompelierung sichtbar
  \includegraphics[height=8.5cm]{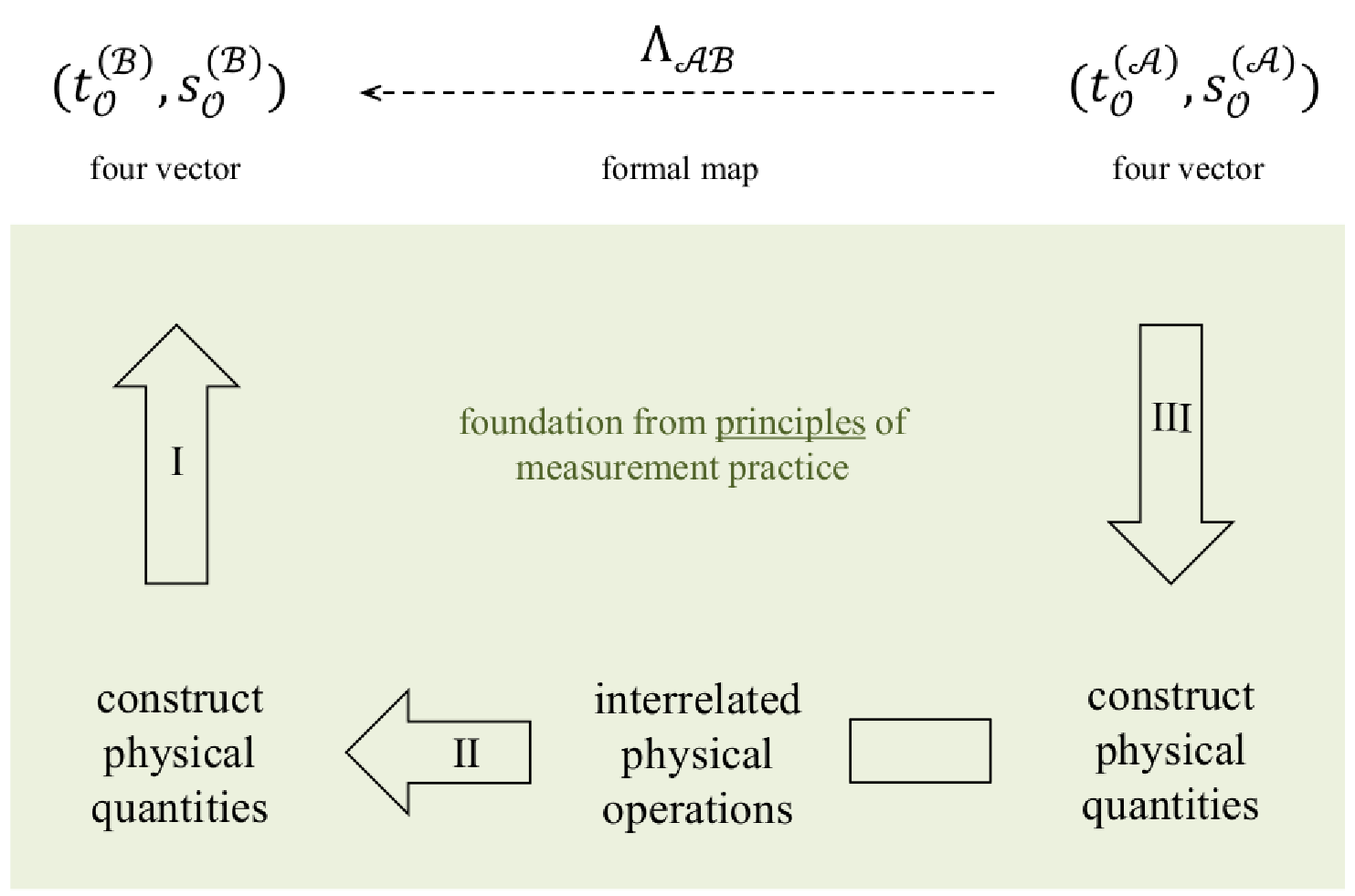}  %\\ if second picture aligned %below...
  \end{center}
  \vspace{-0.0cm}
  \caption{\label{figure-KERN-Bild} physical basis of the mathematical formulation
    }
  \end{figure}

We derive the Lorentz transformation $\Lambda_{\mathcal{A}\mathcal{B}} : (t,s)^{(\mathcal{A})} \mapsto (t,s)^{(\mathcal{B})}$ from $\mathcal{A}$lice and $\mathcal{B}$ob's intrinsic construction of physical quantities of $\mathcal{O}$tto's motion (step I and III) and from their overlapping laser ranging operations (step II) (see figure \ref{figure-KERN-Bild}).
While the formal approach assumes vectors, Lorentz symmetry for its description; we form the mathematical framework. From simple measurement-methodical principles - without mathematical presuppositions - we generate physical quantities  $\left(t^{(\mathcal{A})}_{\overline{\mathcal{P}\mathcal{O}}} \,,\: s^{(\mathcal{A})}_{\overline{\mathcal{P}\mathcal{O}}}\right)$. They specify the layout and number of building blocks in the material model $\mathcal{L}\ast\ldots\ast\mathcal{L}$ which $\mathcal{A}$lice assembles to cover the ''duration'' and ''length'' of measurement object $\overline{\mathcal{P}\mathcal{O}}$. From the interrelation of the underlying practical operations we derive the ''local Lorentz symmetry''.

From classical measurement practice we get Galilei kinematics. Einstein analyzed mutual measurements of moving objects and recognized the need to establish a physical connection between clocks at different locations and speeds. Intrinsic operations with light clocks represent the classical metric locally (Euclidean geometry). For their connection Einstein chose the universal motion of light. By including the (local) light principle and the relativity principle we derive Poincare kinematics. Our locally regular composable grid of light clocks can potentially grow into every direction. It is our \emph{metric connection} between distant measurements. Then the formerly isolated and local notions of the classical metric (absolute time, space, local flatness etc.) will reveal new intricate interrelations.

%%%%%%%%%%%%%%%%%%%%%%%%%%%%%%%%%%%%%%%%%%%%%%%%%%%%%%%%%%%%%%%%%%%%%%%%%

\section{Twin paradox}\label{Kap - Twinparadox}

\begin{figure}         %Bem.: height skaliert Dateibild auf Zielgroesse im Dokument
  \begin{center}         %eps Dateien sind einfuegbar und auch in dvi sichtbar
                         %pdf nur bei direkter pdf-Kompelierung sichtbar
  \includegraphics[height=18cm]{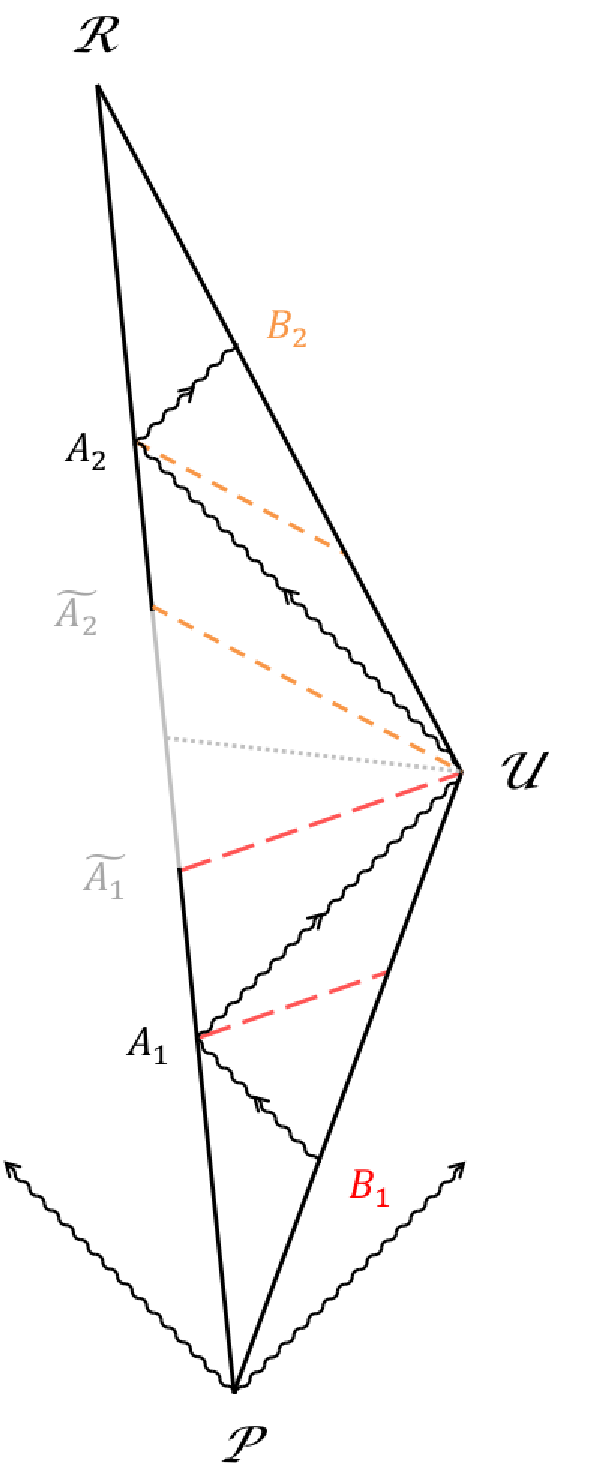}  %\\ if second picture aligned %below...
  \end{center}
  \vspace{-0.0cm}
  \caption{\label{figure-6} connected laser rangings in the Twin configuration of $\mathcal{A}$lice and $\mathcal{B}$ob$_{1,2}$
    }
  \end{figure}

In the Twin configuration $\mathcal{A}$lice and $\mathcal{B}$ob explore their mutual time dilation in a round trip experiment. They depart at moment $\mathcal{P}$. While $\mathcal{A}$lice remains at rest $\mathcal{B}$ob rides with uniform motion $v^{(\mathcal{A})}_{\mathcal{B}}$ to a distant turning point $\mathcal{U}$ and returns with same velocity $-v^{(\mathcal{A})}_{\mathcal{B}}$ to reunite with $\mathcal{A}$lice in future moment $\mathcal{R}$ (see figure \ref{figure-6}).

Throughout the whole round trip $\mathcal{A}$lice can observe $\mathcal{B}$ob. Vice versa $\mathcal{B}$ob will receive all light signals which $\mathcal{A}$lice sends from departure until return. The time during which $\mathcal{A}$lice measures $\mathcal{B}$ob's journey $t^{(\mathcal{A})}_{\mathcal{B}} \equiv t^{(\mathcal{A})}_{\mathcal{A}}$ coincides with her (proper) waiting time; then $\mathcal{B}$ob spends less time on tour $t^{(\mathcal{B})}_{\mathcal{B}} = \underbrace{\sqrt{1-\frac{v_{\mathcal{B}}^2}{c^2}}}_{<1} \; \cdot \; t^{(\mathcal{A})}_{\mathcal{A}}$
than $\mathcal{A}$lice waiting and watching (the previous measurement object $\overline{\mathcal{P}\mathcal{O}}\equiv\mathcal{B}$ coincides with $\mathcal{B}$ob's ride $(t,s)^{(\mathcal{B})}_{\mathcal{B}} \stackrel{(\ref{Formel - Masszsh - Lorentz Trafo st matrix})}{=} \left(\sqrt{1-\frac{v_{\mathcal{B}}^2}{c^2}} \cdot t^{(\mathcal{A})}_{\mathcal{B}}\, , \: 0 \: \right)$). She observes her own clock ticks faster than the moving clock of $\mathcal{B}$ob and vice versa by the symmetry of their relative motion $v^{(\mathcal{B})}_{\mathcal{A}} = - v^{(\mathcal{A})}_{\mathcal{B}}$. During his trip $\mathcal{B}$ob can see all of $\mathcal{A}$lice; if we assume his \emph{observation} time $t^{(\mathcal{B})}_{\mathcal{B}}$
\be
   t^{(\mathcal{B})}_{\mathcal{B}} \stackrel{?}{=} t^{(\mathcal{B})}_{\mathcal{A}} \label{Formel - twins - Bob Beobachtungszeit Wartezeit}
\ee
is the time to \emph{measure} all of $\mathcal{A}$lice $t^{(\mathcal{B})}_{\mathcal{A}}$, then she spends less time waiting  $t^{(\mathcal{A})}_{\mathcal{A}} \! \stackrel{(\ref{Formel - twins - Bob Beobachtungszeit Wartezeit})}{=} \! \underbrace{\sqrt{1-\frac{v_{\mathcal{B}}^2}{c^2}}}_{<1} \; \cdot \; t^{(\mathcal{B})}_{\mathcal{B}}$ than the ride takes for $\mathcal{B}$ob himself (using the reverse Lorentz transformation of $\mathcal{A}$lice waiting interval
$(t,s)^{(\mathcal{A})}_{\mathcal{A}} \stackrel{(\ref{Formel - Masszsh - Lorentz Trafo st matrix})}{=} \left( \sqrt{1-\frac{v_{\mathcal{B}}^2}{c^2}} \; \cdot \; t^{(\mathcal{B})}_{\mathcal{A}} \, , \: 0 \: \right)$). The combined \emph{conclusion} is a contradiction $t^{(\mathcal{B})}_{\mathcal{B}} < t^{(\mathcal{A})}_{\mathcal{A}} < t^{(\mathcal{B})}_{\mathcal{B}}$; the so called \emph{Twin paradox}.

The assumption (\ref{Formel - twins - Bob Beobachtungszeit Wartezeit}) was incorrect; $\mathcal{B}$ob's observation period $t^{(\mathcal{B})}_{\mathcal{B}} \neq t^{(\mathcal{B})}_{\mathcal{A}}$ is not the time for measuring $\mathcal{A}$lice. The Lorentz transformation (\ref{Formel - Masszsh - Lorentz Trafo st matrix}) between physical measures does not refer to durations of observations; it refers to durations of their measurements. For a physically meaningful calculation we remember that basic physical quantities $(s,t)$ originate from tangible operations. The implicit conditions (for constructing the underlying grid of light clock units) must be fulfilled \emph{before} one can apply the Lorentz transformation between results of supposed measurements. Our completed view prevents unreflective calculations in the Lorentz formalism and provides an explanation of the apparent Twin paradox from a measurement-methodical perspective.

$\mathcal{A}$lice can observe and measure $\mathcal{B}$ob throughout the whole trip, unlike $\mathcal{B}$ob. He can receive all light signals from $\mathcal{A}$lice. Though the middle segment of $\mathcal{A}$lice motion $\overline{\widetilde{\mathcal{A}}_1\widetilde{\mathcal{A}}_2}$ is \emph{observable but not measurable} for him. $\mathcal{B}$ob's light clocks $\mathcal{L}^{(\mathcal{B}_1)}$ and $\mathcal{L}^{(\mathcal{B}_2)}$ function properly on the way out and back. Though on each leg $\mathcal{B}$ob cannot cover measurement objects beyond the $\mathcal{U}$-turn point. $\mathcal{B}$ob$_1$'s last indirect laser ranging $\mathcal{B}_1\rightsquigarrow\mathcal{A}_1\rightsquigarrow\widetilde{\mathcal{B}}_1$ reaches $\mathcal{A}$lice at moment $\mathcal{A}_1$ so the reflection (to analyze round trip times) returns before he changes his motion at moment $\mathcal{U}$ (violating measurement condition \{\ref{Kap - SRT Massbestimmung - indirect laser ranging}\}). His direct laser ranging by consecutive and adjacent connection of his light clocks covers $\mathcal{A}$lice up to moment $\widetilde{\mathcal{A}}_1$. Later $\mathcal{B}$ob$_2$ assembles his light clocks $\mathcal{L}^{(\mathcal{B}_2)}$ next to one another forming simultaneity lines with respect to his new state of motion. His direct and indirect laser ranging measurements cover $\mathcal{A}$lice from moment $\widetilde{\mathcal{A}}_2$ resp. $\mathcal{A}_2$ up to the point of $\mathcal{R}$eturn. $\mathcal{B}$ob$_1$ and $\mathcal{B}$ob$_2$ cannot measure segment $\overline{\widetilde{\mathcal{A}}_1\widetilde{\mathcal{A}}_2}$ of $\mathcal{A}$lice motion by intrinsic use of their light clock units.

During his entire round trip $\mathcal{B}$ob measures two segments $\overline{\mathcal{P}\widetilde{\mathcal{A}}_1}$ and $\overline{\widetilde{\mathcal{A}}_2\mathcal{R}}$ of $\mathcal{A}$lice motion. He sees all processes for moving $\mathcal{A}$lice run slower by the same factor $t^{(\mathcal{A})}_{\mathcal{A}} = \sqrt{1-\frac{v_{\mathcal{B}}^2}{c^2}} \; \cdot \; t^{(\mathcal{B})}_{\mathcal{A}}$ as in $\mathcal{A}$lice reverse perspective. Though $\mathcal{B}$ob$_1$ covers a shorter segment of $\mathcal{A}$lice relative motion
\be
   \overline{\mathcal{P}\widetilde{\mathcal{A}}_1} \;\; \sim_{t,s} \;\; \left( \;\; t^{(\mathcal{B}_1)}_{\mathcal{B}_1} \; , \;\; v_{\mathcal{A}} \cdot t^{(\mathcal{B}_1)}_{\mathcal{B}_1} \; \right) \: \cdot \mathcal{L}^{(\mathcal{B}_1)} \nn
\ee
with proper duration $t^{(\mathcal{A})}_{\overline{\mathcal{P}\widetilde{\mathcal{A}}_1}} \stackrel{(\ref{Formel - Masszsh - Lorentz Trafo st matrix})}{=} \sqrt{1-\frac{v_{\mathcal{B}}^2}{c^2}} \cdot t^{(\mathcal{B}_1)}_{\mathcal{B}_1} \stackrel{(\ref{Formel - Masszsh - Lorentz Trafo st matrix})}{=} \left( 1-\frac{v_{\mathcal{B}}^2}{c^2} \right) \cdot \frac{t^{(\mathcal{A})}_{\mathcal{A}}}{2}$. Similarly during his return $\mathcal{B}$ob$_2$ measures a segment of $\mathcal{A}$lice with same proper duration $t^{(\mathcal{A})}_{\overline{\widetilde{\mathcal{A}}_2\mathcal{R}}}$. With regard to measurability by $\mathcal{B}$ob the waiting process of $\mathcal{A}$lice $\overline{\mathcal{P}\mathcal{R}} \equiv \overline{\mathcal{P}\widetilde{\mathcal{A}}_1} \ast \overline{\widetilde{\mathcal{A}}_1\widetilde{\mathcal{A}}_2} \ast \overline{\widetilde{\mathcal{A}}_2\mathcal{R}}$ splits into measurable segments $\overline{\mathcal{P}\widetilde{\mathcal{A}}_1}$ and $\overline{\widetilde{\mathcal{A}}_2\mathcal{R}}$ and non-measurable segment $\overline{\widetilde{\mathcal{A}}_1\widetilde{\mathcal{A}}_2}$. Her waiting time divides accordingly
\bea
   t_{\mathcal{A}} \; & = & \; t_{\overline{\mathcal{P}\widetilde{\mathcal{A}}_1}} \; + \; t_{\overline{\widetilde{\mathcal{A}}_2\mathcal{R}}} \;\;\;\; + \;\;\;\; t_{\overline{\widetilde{\mathcal{A}}_1\widetilde{\mathcal{A}}_2}} \nn \\
   & = & \left( 1-\frac{v_{\mathcal{B}}^2}{c^2} \right) \cdot t_{\mathcal{A}} \;\;\; + \;\;\; \frac{v_{\mathcal{B}}^2}{c^2} \cdot t_{\mathcal{A}} \;\; . \nn
\eea
In the limit where $\mathcal{B}$ob approaches speed of light $v_{\mathcal{B}_i} \rightarrow c$ his total trip duration vanishes
\[
   t_{\mathcal{B}_1} + t_{\mathcal{B}_2} \stackrel{(\ref{Formel - Masszsh - Lorentz Trafo st matrix})}{=} \sqrt{1-\frac{v_{\mathcal{B}}^2}{c^2}}  \cdot t_{\mathcal{A}} \;\; \rightarrow \;\; 0
\]
while during (observable but) unmeasurable part of her waiting process $\mathcal{A}$lice grows older
\[
   t_{\overline{\widetilde{\mathcal{A}}_1\widetilde{\mathcal{A}}_2}} = \frac{v_{\mathcal{B}}^2}{c^2} \cdot t_{\mathcal{A}} \;\; \rightarrow \;\; t_{\mathcal{A}} \;\;\; .
\]

While the formal explanation of the Twin paradox assumes four-vectors $x^{\mu}$, Minkowski metric $\eta_{\mu\nu}$ and ''integrates proper time along a curved worldline'' our measurement-methodical foundation reveals the physical reason and it derives the rules of the calculus as well.

\section{Mathematical formulation of physical operations}\label{Kap - math formulation of physical operations}

We present a novel approach to the foundation of the physical theory, which begins with questions on measurement practice. We are long familiar with basic measurements, as in the case of ''known, very old procedure of length measurements by repeated placement of unit sticks one after the other'' \cite{Carnap Physikalische Begriffsbildung}. Wallot \cite{Wallot - Groessengleichungen Einheiten und Dimensionen} defines ''physical measures are never tangible things, but always \emph{attributes of things}, properties, which we can notice on the things of our experience''. Helmholtz regards attributes of objects, which in a comparison allow the difference of larger, equal or smaller. We want to express their value also numerically (how many times more). The procedure for finding these values is the measurement.

Helmholtz \cite{Helmholtz - Zaehlen und Messen} starts from counting same objects. He begins with fundamental questions about two aspects of basic measurement operations:
\begin{enumerate}
  \item ''What is the physical meaning if we declare two objects as \emph{equal} in a certain relation?''
  \item ''Which character must the physical concatenation of two objects have, that we may consider comparable attributes thereof as connected \emph{additively}?''
\end{enumerate}
In this way Helmholtz elucidates familiar examples like the weight $m_{\mathcal{O}}$ of a material $\mathcal{O}$bject, the length $s_{\overline{\mathcal{A}\mathcal{B}}}$ of a straight line $\overline{\mathcal{A}\mathcal{B}}$, the duration $t_{\mathcal{P}}$ of a physical $\mathcal{P}$rocess etc. Thereby we notice, that the way of concatenation generally depends on the kind of measure. ''We add e.g. weights, by simply placing them on the same weighing-pan. We add time periods, by letting the second begin at exactly the moment, where the first stops; we add lengths, by placing them next to each other in a certain way, namely in a straight line etc.'' A basic measurement therefore requires knowledge of the method of comparison (of a particular attribute of both bodies) ''$\sim_{m}$'' and of the method of their physical concatenation ''$\ast_{m}$''.\footnote{We define derived quantities by equations. Helmholtz calls them more accurately coefficients. ''\emph{Basic quantities} cannot be deduced by equations onto other already explained quantities.''}
\\

We have applied Helmholtz program of direct measurements to relativistic kinematics. We can compare the spatiotemporal order of two objects by the classical probe, whether one of them covers the other. Without one word of mathematics one can manufacture identically constituted light clocks $\mathcal{L}$ and place them literally side by side or one after the other. $\mathcal{A}$lice concatenates these measurement units by a physical process, by letting their inner light rays overlap. Her measurements are based on light principle, classical construction of light clocks $\mathcal{L}$ and their direct and indirect connection by the independent motion of light.

Mathematics comes into being at the moment we introduce units and count, how many congruent building blocks it takes for assembling a regular grid $\mathcal{L} \ast \dots \ast \mathcal{L} \sim_{s} \mathcal{O}$ which covers the measurement object. If both are interchangeable in the comparison they have same length (up to a non-vanishing but practically admissible \emph{measurement error})
\[
   s_{\mathcal{O}} \;\; = \;\;  s_{\mathcal{L} \ast \dots \ast \mathcal{L}} \; + \; \Delta s  \;\; .
\]
\begin{rem}\label{Rem - SRT Kin - doubling of physical measures}
Helmholtz way of basic measurement involves a pair comparison. Measurement object and material model are natural objects. $\mathcal{A}$lice covers the spatiotemporal interval of e.g. relative motion of $\mathcal{B}$ob by a regular grid of ticking light clocks. It is built of solely congruent building blocks $\mathcal{L}$ and it is (locally) invariant under permuting their order; by counting them $\mathcal{A}$lice finds ''how many times'' further and longer $\mathcal{B}$ob's relative motion spreads than the (universal) light in her reference device.
\end{rem}
$\mathcal{A}$lice builds wide layouts of (ticking) light clocks $\mathcal{L} \ast_s \ldots \ast_s \mathcal{L}$ and enduring sequences of (light clock) ticks $\mathcal{L} \ast_t \ldots \ast_t \mathcal{L}$. All steps of her procedure (assembling the material model and conducting length comparison $\sim_s$) are reproducible by any other physicist -- the practical purpose of standardizing measurements (of the magnitude of durations and lengths).\footnote{Her technique is preserved until in empirical practice one oversteps unforeseen conditions, physically specifies them further and thus evolves - in a continual historic process - measurement practice and its (physically determined) mathematical formulation. -- Remembering Feynman's motto: \emph{Yesterday's sensation is today's calibration and tomorrow's background}.}

In starting figure \ref{figure-1} we illustrate only observed objects and observers in motion. Provided the construction of light clocks (measurement unit $\mathcal{L}$) and their connection in consecutive and adjacent ways (concatenation $\ast_t$, $\ast_s$) we introduce increasingly complex material models which where not yet assembled in the uncultivated beginning. As a colloquial expression we introduce \emph{measurement termini}. Along figures \ref{figure-2}, \ref{figure-3}, \ref{figure-4} we define \emph{physical notions} based on measurement operations with light clocks (simultaneous straight measurement path, spatial and temporal projection etc.). Step by step we introduce operational denominations which specify aspects of $\mathcal{A}$lice measurement practice precisely.

From the underlying operational definitions their interrelation becomes transparent. Their common origin inherits a genetic interrelation between measurement termini and the corresponding terms in the mathematical formulation. Based on measurement-methodical principles of their formation we can avoid apparent paradoxes in blind calculations. In retrospect of practical operations with light clocks we emphasize, that all our assertions on the one-way propagation of light strictly come from closed two-way cycles. In laser ranging configuration $\mathcal{A}_1\!\rightsquigarrow\mathcal{B}\rightsquigarrow\mathcal{A}_2$ $\mathcal{A}$lice has no way to measure whether the departing light ray $\overline{\mathcal{A}_1\mathcal{B}}$ towards $\mathcal{B}$ob takes more time than the returning ray $\overline{\mathcal{B}\mathcal{A}_2}$. Similarly we cannot figure what happens \emph{inside} the measurement unit $\mathcal{L} : \mathcal{L}_I\rightsquigarrow \mathcal{L}_{II} \rightsquigarrow \mathcal{L}_I \ldots$ (when light travels between both mirrors to the right $\mathcal{L}_I\rightsquigarrow \mathcal{L}_{II}$ vs. to the left $\mathcal{L}_{II}\rightsquigarrow \mathcal{L}_{I}$).

Direct and indirect laser ranging with light clocks $\mathcal{L}$ always involves both, outgoing and returning pulse. In practice we solely deal with two-way light cycles: (i) inside individual light clocks $\mathcal{L}$ and (ii) in suitably connected configurations of light clocks. Our measurement unit $\mathcal{L}$ comprises both dimensions, width $s_{\mathcal{L}}$ and duration $t_{\mathcal{L}}$ of an elementary two-way light cycle. For the measurement we generate complex configurations of two-way light cycles in suitable layouts of ticking light clocks $\mathcal{L} \ast_t \ldots \ast_t \mathcal{L} \ast_s \ldots \ast_s \mathcal{L}$. Thus $\mathcal{A}$lice covers one-way light rays $\overline{\mathcal{A}_1\mathcal{B}}$ and $\overline{\mathcal{B}\mathcal{A}_2}$ by a layout of two-way light cycles (in light clock grid figure \ref{figure-3}). By counting ticks along her waiting interval $\overline{\mathcal{A}_1\mathcal{A}_2}$ and the light clocks sitting side by side to cover all of laser ranging path $\overline{\mathcal{A}\mathcal{B}}$ $\mathcal{A}$lice measures the magnitude of their length and duration
$ (t,s)_{\overline{\mathcal{A}_1\mathcal{B}}} \stackrel{(\ref{Formel - radar indirekt spatiotemporal physical measure})}{=} \frac{1}{2} \cdot t_{\overline{\mathcal{A}_1\mathcal{A}_2}} + \frac{c}{2} \cdot t_{\overline{\mathcal{A}_1\mathcal{A}_2}}$ resp. $ (t,s)_{\overline{\mathcal{B}\mathcal{A}_2}} = \frac{1}{2} \cdot t_{\overline{\mathcal{A}_1\mathcal{A}_2}} - \frac{c}{2} \cdot t_{\overline{\mathcal{A}_1\mathcal{A}_2}}$.
%[Abschluss: B.H. die Idee Grundmessung mit Mess-EINHEITEN]------------------
\begin{rem}\label{Rem - SRT Kin - inseparable unit}
The meaning of basic physical quantities arises - not by chopping measurement units $\mathcal{L}$ into pieces but instead - by concatenating many congruent measurement units $\mathcal{L}$ (each taken as inseparable unity) to construct material models $\mathcal{L} \ast \dots \ast \mathcal{L}$.
\end{rem}
We introduce all arithmetic operations between measures ''$+$'', ''$-$'', ''$\frac{1}{2} \; \cdot$'' etc. via underlying connection of congruent light clocks. Basic physical quantities specify a reproducible layout of reference devices which covers the measurement objects sufficiently precise.
\\

Our objective is a definition of basic observables from physical operations (what one does in measurement practice). \emph{In absence} of interactions we have developed Helmholtz method for basic measurements of relativistic motion. In this approach, which derives the mathematical formalism of kinematics from this operationalization of length and duration, one can address scope and limitations of the formalism. It can be taken as a basis for our next step, basic measurements of interactions. Next we develop Helmholtz method for the foundation of classical \cite{Hartmann-KM_Dyn} and relativistic dynamics \cite{Hartmann-SRT_Dyn}.
\\
%%%%%%%%%%%%%%%%%%%%%%%%%%%%%%%%%%%%%%%%%%%%%%%%%%%%%%%%%%%%%%%%%%%%%%%
\\
\textbf{Acknowledgements} Thank you to Bruno Hartmann sen. and Peter Ruben for introducing the research problem and essential suggestions. I also want to thank Thomas Thiemann for stimulating discussions and support and Oliver Schlaudt for orientation. This work was made possible initially by the German National Merit Foundation and finally with support by the Perimeter Institute.

%%%%%%%%%%%%%%%%%%%%%%%%%%%%%%%%%%%%%%%%%%%%%%%%%%%%%%%%%%%%%%%%%%%%%%%%%

%\pagebreak

\section*{Appendix A: Successive substitution}\label{Appendix - Rechnungsdetails}

In step I of our series of substitutions we express the space and time component of $\mathcal{B}$ob's physical measure
$\left( \; t^{(\mathcal{B})}_{\overline{\mathcal{P}\mathcal{O}}} \; , \; s^{(\mathcal{B})}_{\overline{\mathcal{P}\mathcal{O}}} \; \right)$
in terms of his laser ranging duration measurements $M$, $N$
\bea
   t^{(\mathcal{B})}_{\overline{\mathcal{P}\mathcal{O}}}
   & \stackrel{(\ref{Formel - Masszsh - Substitution st MN i})}{=} & M + N \nn \\
   s^{(\mathcal{B})}_{\overline{\mathcal{P}\mathcal{O}}}
   & \stackrel{(\ref{Formel - Masszsh - Substitution st MN ii})}{=} & c \cdot N \nn \;\;\; .
\eea
In step II we substitute $\mathcal{B}$ob's laser ranging durations $M$, $N$ - due to the interrelation of their measurement conditions - with $\mathcal{A}$lice laser ranging durations $A$, $B$, $C$
\bea
   M & \stackrel{(\ref{Formel - Masszsh - Substitution MN ABC ii})}{=} & \sqrt{A\cdot(A+B+B)} \label{Formel - Masszsh - Substitution MN ABC - N explizit i} \\
   N & \stackrel{(\ref{Formel - Masszsh - Substitution MN ABC i})(\ref{Formel - Masszsh - Substitution MN ABC ii})}{=} & \frac{1}{2} \cdot \sqrt{A\cdot(A+B+B)} \cdot \frac{A+C+C}{A+B+B} \;\; - \;\; \frac{1}{2} \cdot \sqrt{A\cdot(A+B+B)}
    \label{Formel - Masszsh - Substitution MN ABC - N explizit ii} \;\;\; .
\eea
In step III finally we reformulate $\mathcal{A}$lice laser ranging durations $A$, $B$, $C$ in terms of the space and time components of $\mathcal{A}$lice's physical measures $\left( \; t^{(\mathcal{A})}_{\overline{\mathcal{P}\mathcal{O}}} \; , \; s^{(\mathcal{A})}_{\overline{\mathcal{P}\mathcal{O}}} \; \right)$ and $\left( \; t^{(\mathcal{A})}_{\overline{\mathcal{P}\mathcal{B}}} \; , \; s^{(\mathcal{A})}_{\overline{\mathcal{P}\mathcal{B}}} \; \right)$. We successively insert all substitutions for the space and time component separately
\bea
   s^{(\mathcal{B})}_{\overline{\mathcal{P}\mathcal{O}}} \!\!
   & \stackrel{(\ref{Formel - Masszsh - Substitution st MN ii})}{=} & c \cdot N \nn \\
   & \stackrel{(\ref{Formel - Masszsh - Substitution MN ABC - N explizit ii})}{=} & c \cdot \left[ \frac{1}{2} \cdot \sqrt{A\cdot(A+B+B)} \cdot \frac{A+C+C}{A+B+B} \; - \; \frac{1}{2} \cdot \sqrt{A\cdot(A+B+B)} \cdot \underbrace{\frac{\sqrt{A+B+B}}{\sqrt{A+B+B}}}_{=1} \; \right]
   \nn
\eea
\bea
   & = & c \cdot \frac{1}{2} \cdot \frac{\sqrt{A}}{\sqrt{A+B+B}} \cdot (A+C+C) \; - \; c \cdot \frac{1}{2} \cdot \frac{\sqrt{A}}{\sqrt{A+B+B}} \cdot (A+B+B) \nn \\
   & = & c \cdot \underbrace{\frac{\sqrt{A}}{\sqrt{A}}}_{=1} \cdot \frac{\sqrt{A}}{\sqrt{A+B+B}} \cdot (C-B) \nn \\
   & = & \frac{1}{\sqrt{A\cdot(A+B+B)}} \; \cdot \; \left(\; -\; c \cdot A \cdot B \; + \; c \cdot A \cdot C\right) \nn \\
   & = & \frac{1}{\sqrt{A\cdot(A+B+B)}} \; \cdot \left( \; -\; c \cdot A \cdot B \; \underbrace{ -\; c \cdot B \cdot C \; + \; c \cdot B \cdot C}_{=0} \; + \; c \cdot A \cdot C \right) \nn \\
   & = & \frac{1}{\sqrt{A\cdot(A+B+B)}} \; \cdot \; \left( \; -\; c \cdot B \cdot ( A + C ) \;\; + \;\; c \cdot C \cdot ( A + B ) \right) \nn \\
   & \stackrel{(\ref{Formel - Masszsh - Substitution ABC st i})-(\ref{Formel - Masszsh - Substitution ABC st iv})}{=} & \frac{1}{\sqrt{ \left( t^{(\mathcal{A})}_{\overline{\mathcal{P}\mathcal{B}}} \; - \; \frac{1}{c} \cdot s^{(\mathcal{A})}_{\overline{\mathcal{P}\mathcal{B}}} \right) \cdot \left( t^{(\mathcal{A})}_{\overline{\mathcal{P}\mathcal{B}}} \; + \; \frac{1}{c} \cdot s^{(\mathcal{A})}_{\overline{\mathcal{P}\mathcal{B}}} \right) }} \cdot  \left( - s^{(\mathcal{A})}_{\overline{\mathcal{P}\mathcal{B}}} \cdot t^{(\mathcal{A})}_{\overline{\mathcal{P}\mathcal{O}}} \; + \; s^{(\mathcal{A})}_{\overline{\mathcal{P}\mathcal{O}}} \cdot t^{(\mathcal{A})}_{\overline{\mathcal{P}\mathcal{B}}}  \right) \nn \\
   & = & \frac{t_{\overline{\mathcal{P}\mathcal{B}}}}{\sqrt{  {t_{\overline{\mathcal{P}\mathcal{B}}}}^2 \; - \; \frac{1}{c^2} \cdot {s_{\overline{\mathcal{P}\mathcal{B}}}}^2  }} \; \cdot \; \left( - \frac{s_{\overline{\mathcal{P}\mathcal{B}}}}{t_{\overline{\mathcal{P}\mathcal{B}}}} \cdot t_{\overline{\mathcal{P}\mathcal{O}}} \; + \; s_{\overline{\mathcal{P}\mathcal{O}}}  \right) \nn \\
   & = & \frac{1}{\sqrt{ \frac{{t_{\overline{\mathcal{P}\mathcal{B}}}}^2}{{t_{\overline{\mathcal{P}\mathcal{B}}}}^2}  \; - \; \frac{1}{c^2} \cdot \frac{{s_{\overline{\mathcal{P}\mathcal{B}}}}^2}{{t_{\overline{\mathcal{P}\mathcal{B}}}}^2}  }} \; \cdot \; \left( - v_{\mathcal{B}} \cdot t_{\overline{\mathcal{P}\mathcal{O}}} \; + \; s_{\overline{\mathcal{P}\mathcal{O}}}  \right) \nn \\
s^{(\mathcal{B})}_{\overline{\mathcal{P}\mathcal{O}}}
   & = &   - \; \frac{1}{\sqrt{ 1  \; - \; \frac{ {v_{\mathcal{B}}}^2 }{c^2} }} \; \cdot \; v_{\mathcal{B}} \cdot \;\; t^{(\mathcal{A})}_{\overline{\mathcal{P}\mathcal{O}}} \;\;\; + \;\;\; \frac{1}{\sqrt{ 1  \; - \; \frac{ {v_{\mathcal{B}}}^2 }{c^2} }} \; \cdot \;\; s^{(\mathcal{A})}_{\overline{\mathcal{P}\mathcal{O}}} \label{Formel - Masszsh - Lorentz Trafo s}
\eea
where $\mathcal{A}$lice has determined the velocity of the relative motion of Bob $v_{\mathcal{B}} := s^{(\mathcal{A})}_{\overline{\mathcal{P}\mathcal{B}}} \!\left/\! t^{(\mathcal{A})}_{\overline{\mathcal{P}\mathcal{B}}}\right.$ and with notation simplified in last steps on $\mathcal{A}$lice right hand side by suppressing her indices$^{(\mathcal{A})}$.
\bea
   \;\;\; t^{(\mathcal{B})}_{\overline{\mathcal{P}\mathcal{O}}} \!\!
   & \stackrel{(\ref{Formel - Masszsh - Substitution st MN i})}{=} & N + M \nn \\
   & \!\!\! \stackrel{(\ref{Formel - Masszsh - Substitution MN ABC - N explizit i})(\ref{Formel - Masszsh - Substitution MN ABC - N explizit ii})}{=} \!\!\! & \left[ \frac{1}{2} \cdot \sqrt{A\cdot(A+B+B)} \cdot \frac{A+C+C}{A+B+B} \; + \; \frac{1}{2} \cdot \sqrt{A\cdot(A+B+B)} \cdot \underbrace{\frac{\sqrt{A+B+B}}{\sqrt{A+B+B}}}_{=1} \; \right]
   \nn \\
   & = & \frac{1}{2} \cdot \frac{\sqrt{A}}{\sqrt{A+B+B}} \cdot (A+C+C) \; + \; \frac{1}{2} \cdot \frac{\sqrt{A}}{\sqrt{A+B+B}} \cdot (A+B+B) \nn \\
   & = & \underbrace{\frac{\sqrt{A}}{\sqrt{A}}}_{=1} \cdot \frac{\sqrt{A}}{\sqrt{A+B+B}} \cdot (A+B+C) \nn
\eea
\bea
   & = & \frac{1}{\sqrt{A\cdot(A+B+B)}} \; \cdot \; \left(\; A \cdot A \; + \; A \cdot B \: + \; A \cdot C \; \underbrace{+ \; B \cdot C \: - \; B \cdot C}_{=0} \right) \nn \\
   & = &  \frac{1}{\sqrt{A\cdot(A+B+B)}} \; \cdot \; \left(\; (A + C) \cdot (A+B) \: - \; B \cdot C \right) \nn \\
   & \!\!\!\!\! \stackrel{(\ref{Formel - Masszsh - Substitution ABC st i})-(\ref{Formel - Masszsh - Substitution ABC st iv})}{=} \!\!\!\!\! & \frac{1}{\sqrt{ \left( t^{(\mathcal{A})}_{\overline{\mathcal{P}\mathcal{B}}} \; - \; \frac{1}{c} \cdot s^{(\mathcal{A})}_{\overline{\mathcal{P}\mathcal{B}}} \right) \cdot \left( t^{(\mathcal{A})}_{\overline{\mathcal{P}\mathcal{B}}} \; + \; \frac{1}{c} \cdot s^{(\mathcal{A})}_{\overline{\mathcal{P}\mathcal{B}}} \right) }} \cdot  \left(  t^{(\mathcal{A})}_{\overline{\mathcal{P}\mathcal{O}}} \cdot t^{(\mathcal{A})}_{\overline{\mathcal{P}\mathcal{B}}} \; - \; \frac{1}{c} \cdot s^{(\mathcal{A})}_{\overline{\mathcal{P}\mathcal{B}}} \cdot  \frac{1}{c} \cdot s^{(\mathcal{A})}_{\overline{\mathcal{P}\mathcal{O}}}  \right) \;\;\;\;\;\;\;\: \nn \\
   & = & \frac{t_{\overline{\mathcal{P}\mathcal{B}}}}{\sqrt{  {t_{\overline{\mathcal{P}\mathcal{B}}}}^2 \; - \; \frac{1}{c^2} \cdot {s_{\overline{\mathcal{P}\mathcal{B}}}}^2  }} \; \cdot \; \left( t_{\overline{\mathcal{P}\mathcal{O}}} \; - \; \frac{1}{c^2} \cdot \frac{s_{\overline{\mathcal{P}\mathcal{B}}}}{t_{\overline{\mathcal{P}\mathcal{B}}}} \cdot
   s_{\overline{\mathcal{P}\mathcal{O}}}  \right) \nn \\
   t^{(\mathcal{B})}_{\overline{\mathcal{P}\mathcal{O}}}
   & = &  \frac{1}{\sqrt{ 1  \; - \; \frac{ {v_{\mathcal{B}}}^2 }{c^2} }} \; \cdot \;\; t^{(\mathcal{A})}_{\overline{\mathcal{P}\mathcal{O}}} \;\;\;\; - \;\;\; \frac{1}{\sqrt{ 1  \; - \; \frac{ {v_{\mathcal{B}}}^2 }{c^2} }} \cdot \frac{v_{\mathcal{B}}}{c^2} \; \cdot
   \;\; s^{(\mathcal{A})}_{\overline{\mathcal{P}\mathcal{O}}} \label{Formel - Masszsh - Lorentz Trafo t}
\eea

%%%%%%%%%%%%%%%%%%%%%%%%%%%%%%%%%%%%%%%%%%%%%%%%%%%%%%%%%%%%%%%%%%%%%%%%

\end{document}